\documentclass[
aps,
twocolumn, 
showpacs,
prl, 
superscriptaddress,
tightenlines,
10pt] 
{revtex4-1}

\usepackage{graphicx}
\usepackage{color}
\usepackage{ulem}
\usepackage{verbatim}
\usepackage{amsmath}
\usepackage{amssymb}
\usepackage[left=1.3cm,right=1.3cm,top=2.2cm,bottom=1.5cm,footskip=1cm]{geometry}

\usepackage{etoolbox}
\usepackage{lipsum}

\usepackage{array}
\usepackage{tabulary}
\newcolumntype{K}[1]{>{\centering\arraybackslash}p{#1}}

\usepackage{hyperref}
\hypersetup{
  colorlinks   = true, 
  urlcolor     = blue, 
  linkcolor    = blue, 
  citecolor    = blue 
}

\begin{document}
\fontsize{9pt}{10.8pt} 

\title{\LARGE Photonic quantum state transfer between a cold atomic gas and a crystal}



\author{Nicolas Maring}
\affiliation{ICFO-Institut de Ciencies Fotoniques, The Barcelona Institute of Science and Technology, 08860 Castelldefels (Barcelona), Spain}

\author{Pau Farrera}
\affiliation{ICFO-Institut de Ciencies Fotoniques, The Barcelona Institute of Science and Technology, 08860 Castelldefels (Barcelona), Spain}

\author{Kutlu Kutluer}
\affiliation{ICFO-Institut de Ciencies Fotoniques, The Barcelona Institute of Science and Technology, 08860 Castelldefels (Barcelona), Spain}

\author{Margherita Mazzera}
\affiliation{ICFO-Institut de Ciencies Fotoniques, The Barcelona Institute of Science and Technology, 08860 Castelldefels (Barcelona), Spain}

\author{Georg Heinze}
\email[Contact: ]{georg.heinze@alumni.icfo.eu}
\affiliation{ICFO-Institut de Ciencies Fotoniques, The Barcelona Institute of Science and Technology, 08860 Castelldefels (Barcelona), Spain}

\author{Hugues de Riedmatten}
\homepage[]{http://qpsa.icfo.es}
\email[Contact: ]{hugues.deriedmatten@icfo.eu}
\affiliation{ICFO-Institut de Ciencies Fotoniques, The Barcelona Institute of Science and Technology, 08860 Castelldefels (Barcelona), Spain}
\affiliation{ICREA-Instituci\'{o} Catalana de Recerca i Estudis Avan\c cats, 08015 Barcelona, Spain}%


\maketitle

\textbf{
Interfacing fundamentally different quantum systems is key to build future hybrid quantum networks \cite{Kimble2008}. Such heterogeneous networks offer superior capabilities compared to their homogeneous counterparts as they merge individual advantages of disparate quantum nodes in a single network architecture \cite{Wallquist2009,Walmsley2016}.
However, only very few investigations on optical hybrid-interconnections have been carried out due to the high fundamental and technological challenges, which involve e.g. wavelength and bandwidth matching of the interfacing photons. Here we report the first optical quantum interconnection between two disparate matter quantum systems with photon storage capabilities. We show that a quantum state can be faithfully transferred between a cold atomic ensemble \cite{Chou2005, Chaneliere2005} and a rare-earth doped crystal \cite{DeRiedmatten2008, Hedges2010, Clausen2011,Saglamyurek2011} via a single photon at telecommunication wavelength, using cascaded quantum frequency conversion. We first demonstrate that quantum correlations between a photon and a single collective spin excitation in the cold atomic ensemble can be transferred onto the solid-state system. We also show that single-photon time-bin qubits generated in the cold atomic ensemble can be converted, stored and retrieved from the crystal with a conditional qubit fidelity of more than $85\%$. Our results open prospects to optically connect quantum nodes with different capabilities and represent an important step towards the realization of large-scale hybrid quantum networks.}\\

With the advent of quantum technologies scientists are now trying to build quantum networks which are expected to be much more powerful than the simple sum of their constituents \cite{Kimble2008}. Pioneering experiments in this line of research include the photonic coupling of identical quantum nodes, such as atomic ensembles \cite{Chou2005, Chaneliere2005, Eisaman2005}, single trapped atoms \cite{Ritter2012} and ions \cite{Moehring2007}, and solid-state devices \cite{Usmani2012,Pfaff2014,Delteil2017}. However, each platform comes along with individual functionalities, e.g., in terms of processing and storage. Hence, a hybrid quantum network, which benefits from the strengths of different platforms, would offer more capabilities than a network consisting of identical quantum systems. Although significant efforts have been devoted to build hybrid quantum systems e.g. devices combining different quantum systems on a single chip \cite{Xiang2012}, or different species of closely spaced trapped ions \cite{Tan2015,Ballance2015},  interactions between these systems are typically mediated by microwave photons or Coulomb interactions, which are not favorable for long distance quantum communication. 

Instead, photonic interconnections between different quantum systems have so far been realized only in very few experiments \cite{Akopian2011, Siyushev2014, Meyer2015, Tang2015}, however neither demonstrating quantum state transfer nor interfacing two different long-lived quantum memory (QM) systems, which are both crucial requirements for quantum networks applications. A photonic quantum interconnection between different platforms was demonstrated in Ref.~\cite{Lettner2011}, using a single atomic species. The main challenge to efficiently interface two different quantum systems via a photonic link, is to obtain strong light-matter interaction between a single mediating photon and both matter systems, whose atomic transitions can differ significantly in wavelength and linewidth. 

In this letter, we demonstrate quantum state transfer between two fundamentally different quantum memory systems via a single photon at telecom wavelength. 
On the one hand, we use a laser cooled ensemble of $^{87}\mathrm{Rb}$ atoms, which, besides being an excellent quantum memory and single photon source \cite{Sangouard2011}, also gives access to tunable non-linear interactions enabling quantum processing via Rydberg excitations \cite{Saffman2010}. On the other hand, we use a rare earth ion doped crystal ($\mathrm{Pr}^{3+}\mathrm{:Y}_2\mathrm{SiO}_5$) exhibiting outstanding properties for multiplexed long-lived quantum state storage \cite{Hedges2010,Ferguson2016,Kutluer2017,Seri2017}. To overcome the wavelength gap between both systems, we use quantum frequency conversion techniques to convert photons emitted by the $^{87}\mathrm{Rb}$ QM from $780\,\mathrm{nm}$ to the telecom C-band at $1552\,\mathrm{nm}$ and then back to $606\,\mathrm{nm}$ to resonantly interact with the $\mathrm{Pr}^{3+}\mathrm{:Y}_2\mathrm{SiO}_5$ crystal.
We show that a single collective spin excitation (spin-wave) stored in the cold atomic QM can be optically transferred onto a long-lived collective optical excitation inside the crystal.
By transmitting correlated single photons and qubits, we demonstrate quantum correlation preserving and coherent quantum state transfer between the disparate quantum nodes.

\begin{figure*}
\includegraphics[width=1.0\textwidth]{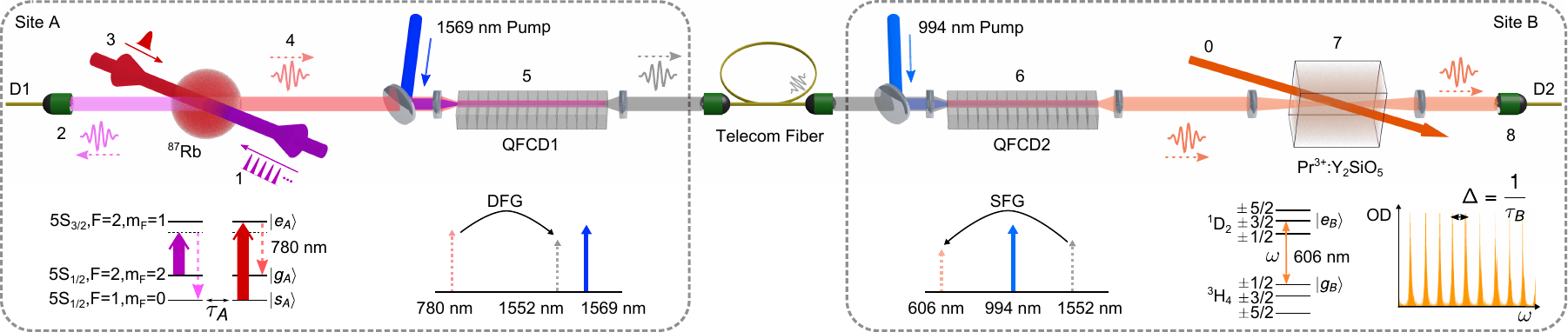}
\caption{\textbf{Schematic setup and relevant level schemes.} At site \textit{A} a cold cloud of $^{87}\mathrm{Rb}$ atoms is held inside a MOT. Following the DLCZ protocol, non-classically correlated photon pairs are produced by first sending classical write pulses (1) generating a spin-wave inside the atomic cloud heralded by a write photon (2) which is spectrally filtered by a monolithic Fabry-Perot cavity (not shown). Upon a write photon detection at D1, the spin-wave is read-out by sending a classical read pulse (3) generating the read photon (4). QFCD1 consists of a periodically poled lithium niobate (PPLN) crystal with an integrated proton exchange (PE) waveguide continuously pumped by a strong pump laser at $1569\,\mathrm{nm}$. It converts the read photon from $780\,\mathrm{nm}$ to $1552\,\mathrm{nm}$ (5). The converted photon is then separated from the strong pump light via dielectric band pass filters (not shown) before it is sent via a telecom fiber to site~\textit{B} where QFCD2 (consisting of a PPLN ridge waveguide pumped by strong $994\,\mathrm{nm}$ laser radiation) converts it to $606\,\mathrm{nm}$ via SFG (6) before the photon is again spectrally filtered by several elements (not shown, see Methods). The $\mathrm{Pr}^{3+}\mathrm{:Y}_{2}\mathrm{SiO}_{5}$ crystal was initially prepared with an AFC (0) using a strong preparation beam at $606\,\mathrm{nm}$, to store the converted read photon (7). After retrieval it is finally detected at D2 (8).}
\label{Figure1}
\end{figure*}

The basic concept of our experiment can be understood along the schematic depicted in Fig.~\ref{Figure1} which is separated into two main sites \textit{A} and \textit{B} (for a detailed figure and description see Methods). 
At site \textit{A}, we operate a $^{87}\mathrm{Rb}$ magneto optical trap (MOT) to generate synchronizable single photons of controllable bandwidth and temporal shape (see Methods) which, afterwards, are frequency converted from $780\,\mathrm{nm}$ to $1552\,\mathrm{nm}$ in an all-solid-state quantum frequency conversion device (QFCD). Following the Duan-Lukin-Cirac-Zoller (DLCZ) protocol \cite{Duan2001}, we send a series of classical write pulses onto the Rb atoms to create Raman-scattered write photons with intrinsic excitation probability $p_\mathrm{e}$ per optical mode and trial. A single optical mode of the isotropically emitted write photons is collected and sent to single photon detector (SPD) D1. A detection event at D1 heralds the generation of a spin-wave involving long-lived ground states of the $^{87}\mathrm{Rb}$ ensemble. After a programmable storage time, we apply a classical read pulse, to deterministically read-out the spin-wave and generate a read photon in a well defined spatio-temporal mode, with a fiber-coupled retrieval efficiency of $\eta_\mathrm{ret}^{A}\approx30\%$. Depending on the temporal shape of the read pulse, we generate single read photons with Gaussian envelope or time-bin envelope exhibiting a sub-natural linewidth of about $2\,\mathrm{MHz}$ (see Methods). 
The read single photon is then sent to the first QFCD, which converts it from $780\,\mathrm{nm}$ to $1552\,\mathrm{nm}$ via difference frequency generation (DFG) \cite{Albrecht2014} with an internal conversion efficiency of $\eta_\mathrm{int}^\mathrm{QFCD1}=56\%$. After noise filtering, the converted photon is then coupled into a $10\,\mathrm{m}$ telecom fiber and sent to site~\textit{B} in another laboratory.

At site~\textit{B}, the telecom read photon is first back-converted to  $606\,\mathrm{nm}$ ($\eta_\mathrm{int}^\mathrm{QFCD2}=60\%$) via sum frequency generation (SFG) in QFCD2. After noise filtering the single photon is sent to the $\mathrm{Pr}^{3+}\mathrm{:Y}_{2}\mathrm{SiO}_{5}$ crystal inside a cryostat at a temperature of $3.5\,\mathrm{K}$. We use the atomic frequency comb (AFC) scheme \cite{DeRiedmatten2008} to store, analyze, and retrieve the converted single photon. We create an AFC of $4\,\mathrm{MHz}$ width with absorption peaks spaced by $\Delta=400\,\mathrm{kHz}$ on the optical transition of $\mathrm{Pr}^{3+}$ at $606\,\mathrm{nm}$. Then, the converted single photon is stored by the AFC and collectively re-emitted with an efficiency of $\eta^{B}=30\%$ after a pre-defined storage time of $\tau_B=1/\Delta=2.5\,\mu\mathrm{s}$ before being detected by SPD~D2. The probability to obtain an emitted, converted, stored and retrieved photon after the crystal, conditioned on a write photon detection at D1, is approximately $10^{-3}$. This includes 1.2$\%$ total conversion efficiency (with all optical losses) from 780 nm to 606 nm.

An important experimental requirement is to precisely match the central frequencies of the converted read photon and the prepared AFC, and to minimize the linewidths of all the lasers involved to ensure efficient storage and stable interference conditions for the qubit analysis. We estimate that the frequency stability of the converted photon needs to be significantly better than $1\,\mathrm{MHz}$ (see Methods). This is done by active frequency stabilization and a chopped beat-note lock between classical $780\,\mathrm{nm}$ light converted in the QFCDs and the $606\,\mathrm{nm}$ preparation laser as reference (see Methods). 

\begin{figure}
\includegraphics[width=1.0\columnwidth]{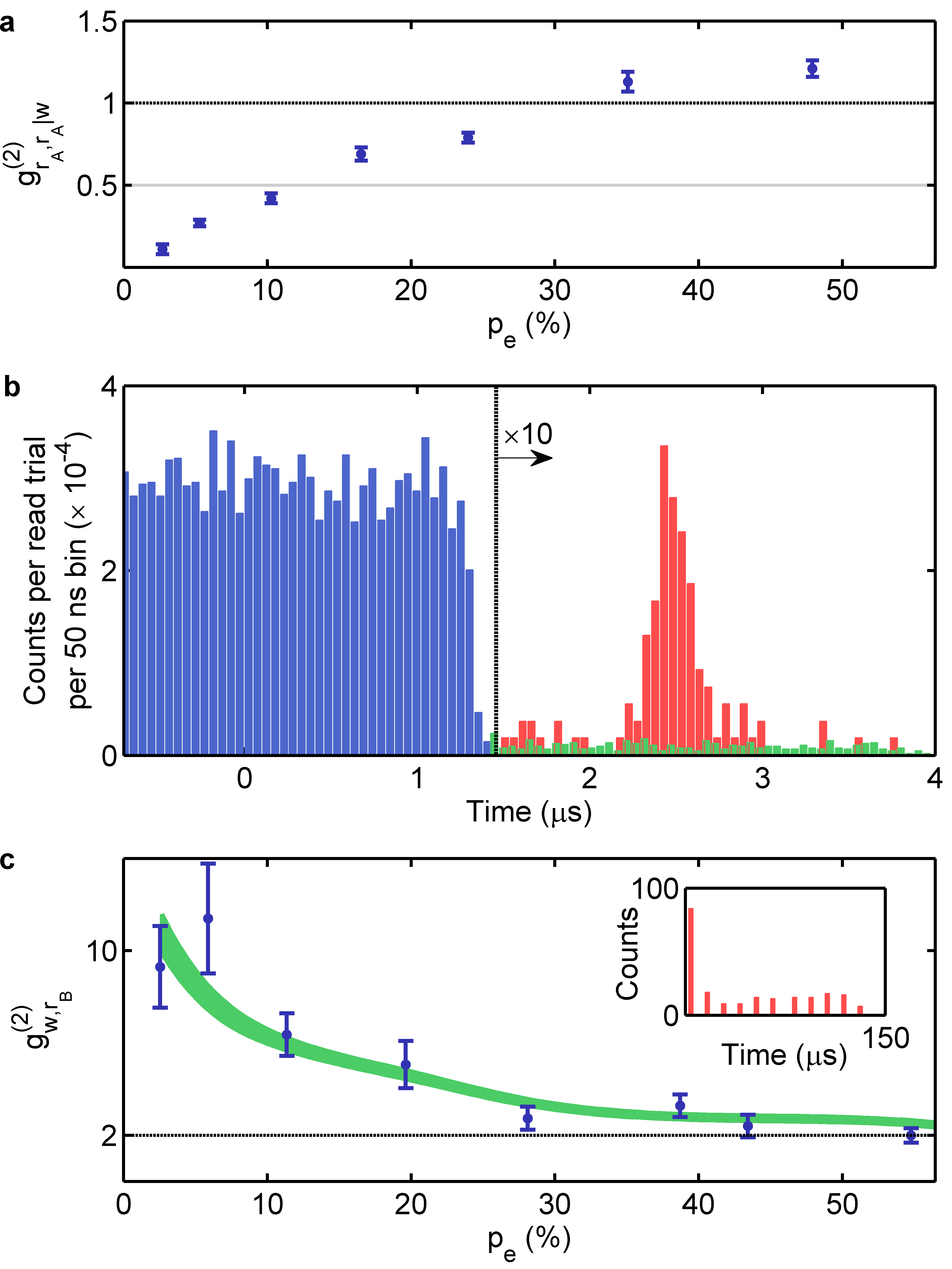}
\caption{\textbf{Photon generation, conversion and storage.} (a) Anti-bunching parameter of the read photon after the MOT vs $p_\mathrm{e}$. The dashed line indicates the threshold for classical states ($\alpha\geq1$) and the dotted line for a two-photon Fock state ($\alpha=0.5$). (b) Time histogram of detections at D2 if the spin-wave (excitation probability $p_\mathrm{e}\approx 35\%$) is read-out from the cold atomic Rb QM, the photons are frequency converted in the QFCDs, and stored at $t=0$ in the crystal. During storage (at $t=1.2\,\mu\mathrm{s}$), the pump of QFCD2 is gated off, and the re-emitted photons are detected as a pronounced AFC echo at $t=2.5\,\mu\mathrm{s}$ (red trace, detected coincidence rate $\sim90\,/\mathrm{h}$ in a $400\,\mathrm{ns}$ window around the echo). The green trace corresponds to the noise level, i.e. if no read photon is sent. (c) Normalized cross-correlation $g^{(2)}_{w,r_B}$ between the write photons from the cold atomic QM and the converted, stored and retrieved read photons from the crystal for different $p_\mathrm{e}$. The green area corresponds to the expected $g^{(2)}_{w,r_B}$ as deduced from a similar model as in \cite{Albrecht2014}. The dashed line represents the classical upper bound $g^{(2)}_{w,r_B} \leq 2$.  The inset shows a typical $g^{(2)}$ histogram of coincidence detections for several read-out trials separated by the trial period of $\sim13\,\mu\mathrm{s}$ obtained at $p_\mathrm{e}\approx11\%$. Error bars correspond to $\pm1\,\mathrm{s.d.}$ of the photon counting statistics.}
\label{Figure2}
\end{figure}

Before interfacing both quantum systems, we characterize the read photons generated at site~\textit{A}. Fig.~\ref{Figure2}(a) shows the heralded autocorrelation function $\alpha=g^{(2)}_{r_A,r_A|w}$ (see Methods) for different $p_\mathrm{e}$ measured via a Hanbury Brown Twiss setup inserted directly after the MOT. We obtain strongly anti-bunched read photons in the single photon regime ($\alpha<0.5$) for low $p_\mathrm{e}\lesssim11\%$, in the non-classical regime ($\alpha<1$) for $p_\mathrm{e}\lesssim25\%$, before surpassing the classical threshold for higher $p_\mathrm{e}$ due to multiple spin-wave excitations.

We now present photon generation, conversion, and storage involving the whole experimental setup. We first verify that photons emitted by the atomic QM can be successfully converted and stored in the crystal. We create at site $A$ a heralded  $200\,\mathrm{ns}$ long (FWHM) Gaussian read photons at $p_\mathrm{e}\approx 35\%$. 
Figure~\ref{Figure2}(b) shows the histogram of detection events at D2. The photons arrive at the crystal at $t=0\,\mu\mathrm{s}$, however, no leakage is visible here, as they are buried in the noise generated by the QFCDs. The noise is suppressed at $t=1.2\,\mu\mathrm{s}$ by gating off the pump of QFCD2. At $t=2.5\,\mu\mathrm{s}$ we detect a pronounced echo signature from the retrieved read photons with a Signal to Noise Ratio $\mathrm{SNR}=17\pm2$, mostly limited by the dark counts of D2. The echo shows the same Gaussian temporal shape as the initial read photons with a FWHM of $200\,\mathrm{ns}$. 

To investigate the non-classicality of the state transfer, we measured the normalized cross-correlation function $g^{(2)}_{w,r_B}$ (see Methods) of the converted, stored and retrieved photons with the initial write photons for different $p_\mathrm{e}$ by comparing coincidences in different storage trials (see Fig.~\ref{Figure2}(c)). At $p_\mathrm{e} \approx 5\%$ (with a coincidence rate of approximately 20 counts per hour) we obtain $g^{(2)}_{w,r_B} = 11.4\pm 2.4$, demonstrating quantum-correlation preserving state-transfer, as the $g^{(2)}_{w,r_B}$ value stays well above the classical bound of $g^{(2)}=2$ assuming thermal statistics for the write and read photons (see Methods). The same holds true for a broad range of $p_\mathrm{e}$ and for storage times $\tau_B$ up to $10\,\mu\mathrm{s}$ (see Methods). The experimental data in Fig.~\ref{Figure2}(c) match well with the expected behavior (green area) calculated via a simple model taking into account the measured cross-correlation $g^{(2)}_{w,r_A}$ after the MOT and the total SNR of the read photon after conversion and storage \cite{Albrecht2014}.

\begin{figure}
\includegraphics[width=1.0\columnwidth]{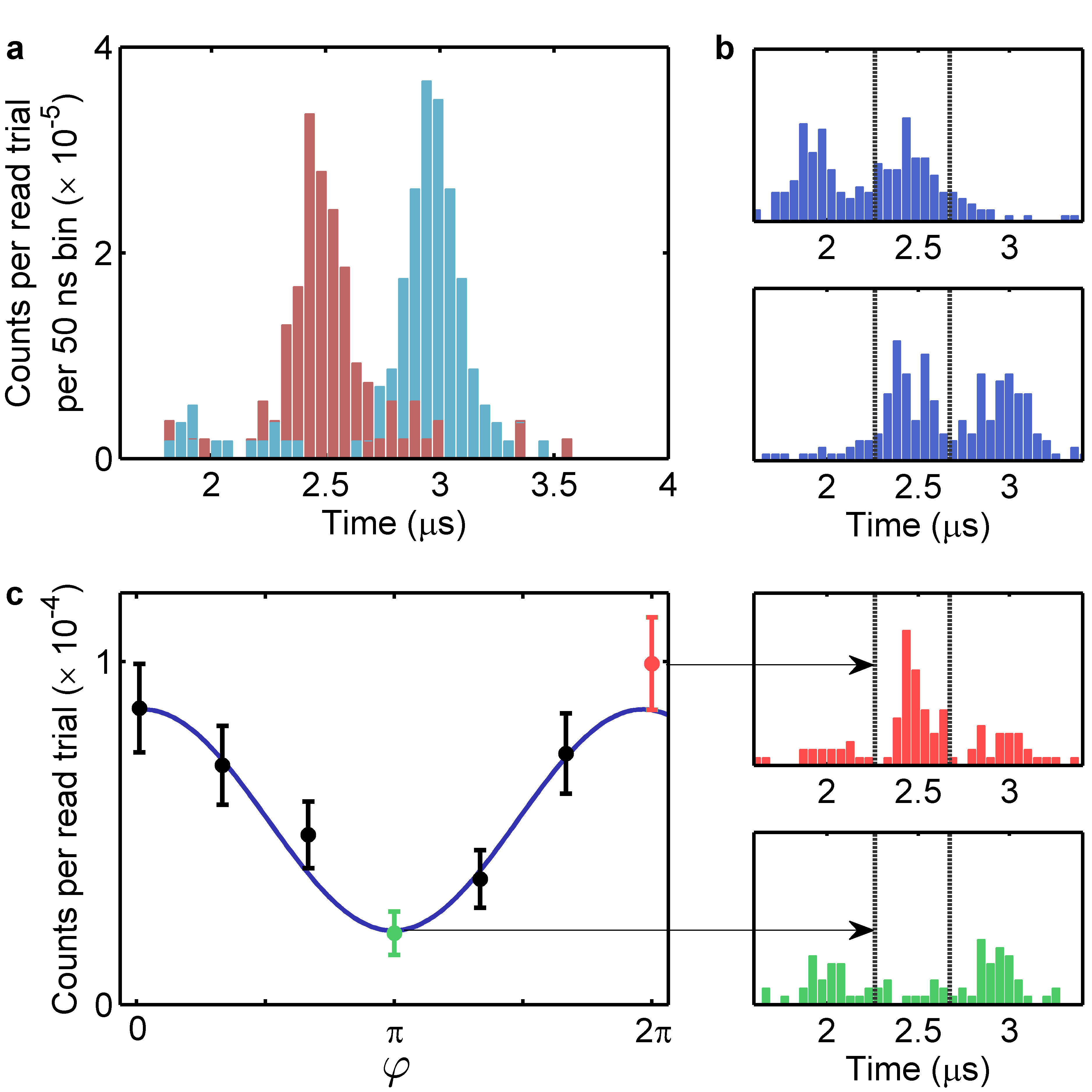}
\caption{\textbf{Coherence Preservation.} (a) Detected echo of Gaussian shaped read photons which were created ($p_\mathrm{e}\approx35\%$) at either an early time $t=0$ (red) or at a later time $t=0.5\,\mu\mathrm{s}$ (blue) at site~\textit{A} and stored for $\tau_{B}=2.5\,\mu\mathrm{s}$ at site~\textit{B}. (b) Time-bin photons stored and retrieved after either a short storage time $\tau_{B_1}=2\,\mu\mathrm{s}$ (top) or a long one $\tau_{B_2}=2.5\,\mu\mathrm{s}$ (bottom). (c) Time-bin interference fringe, i.e. coincidence counts between initial write photon detections at D1 and detection events during the time-bin-overlap at D2 if a time-bin read photon is stored and retrieved from the crystal prepared with two AFCs. Error bars correspond to $\pm1\,\mathrm{s.d.}$ of the photon counting statistics. On the right two examples of time histograms between detection events at D1 and D2 are shown for $\varphi=0^{\circ}$ (top) and $\varphi=180^{\circ}$ (bottom). The $400\,\mathrm{ns}$ coincidence window where the time-bins overlap is indicated by dashed lines.}
\label{Figure3}
\end{figure}

Next, we studied the coherence properties of the state transfer between the two different quantum systems. We use time-bin qubits, which offer advantages for long distance quantum communication due to their robustness against external perturbations. If a heralding write photon is detected at D1, we shape the read pulse in a way that the spin-wave stored in the Rb QM is mapped onto a photonic time-bin qubit $|\Psi\rangle_A=c_1|e\rangle+c_2e^{i\varphi}|l\rangle$, where $|e\rangle$ and $|l\rangle$ represent early and late time-bin, $\varphi$ is their relative phase, controlled by the phase of the second read pulse, and $c_1^2+c_2^2=1$ (see Methods). To store the photonic time-bin qubit, we take advantage of the intrinsic temporal multimodality of the AFC scheme \cite{DeRiedmatten2008}.

Figure~\ref{Figure3}(a) shows the time histogram of detection events at D2 of the early and late time-bins (created at $p_\mathrm{e}\approx35\%$) sent through the QFCDs, and stored and retrieved from the crystal prepared with a single AFC of $\tau_B=2.5\,\mu\mathrm{s}$ storage time. The two echoes represent the polar states of a time-bin qubit and exhibit an average SNR above $19 \pm 2$. If a delocalized time-bin photon ($c_1=c_2=\frac{1}{\sqrt{2}}$) is created in the Rb QM, converted in the QFCDs and stored in the crystal for either $\tau_{B_1}=2\,\mu\mathrm{s}$ or $\tau_{B_2}=2.5\,\mu\mathrm{s}$, we detect the histograms shown in Fig.~\ref{Figure3}(b). 
In order to analyze the qubit, we use the crystal as an interferometer by preparing two overlapped AFCs with storage times $\tau_{B_1}$ and $\tau_{B_2}$ ($\eta^B = 10\%$ each) \cite{DeRiedmatten2008}. In that case, we obtain the histograms shown in the right panels of Fig.~\ref{Figure3}(c). These two histograms were recorded with a phase shift of $\varphi=0^{\circ}$ (top) and $\varphi=180^{\circ}$ (bottom) between the early and late time-bin. Strong interference between the two temporal modes can be seen in the central region where the time-bins overlap. Measuring the coincidences in that time window vs. $\varphi$ gives the interference fringe depicted in Fig.~\ref{Figure3}(c) with a fitted visibility of $V=60\pm9.9\%$, confirming the high degree of coherence preservation between the two disparate quantum systems.

\begin{figure}
\includegraphics[width=1.0\columnwidth]{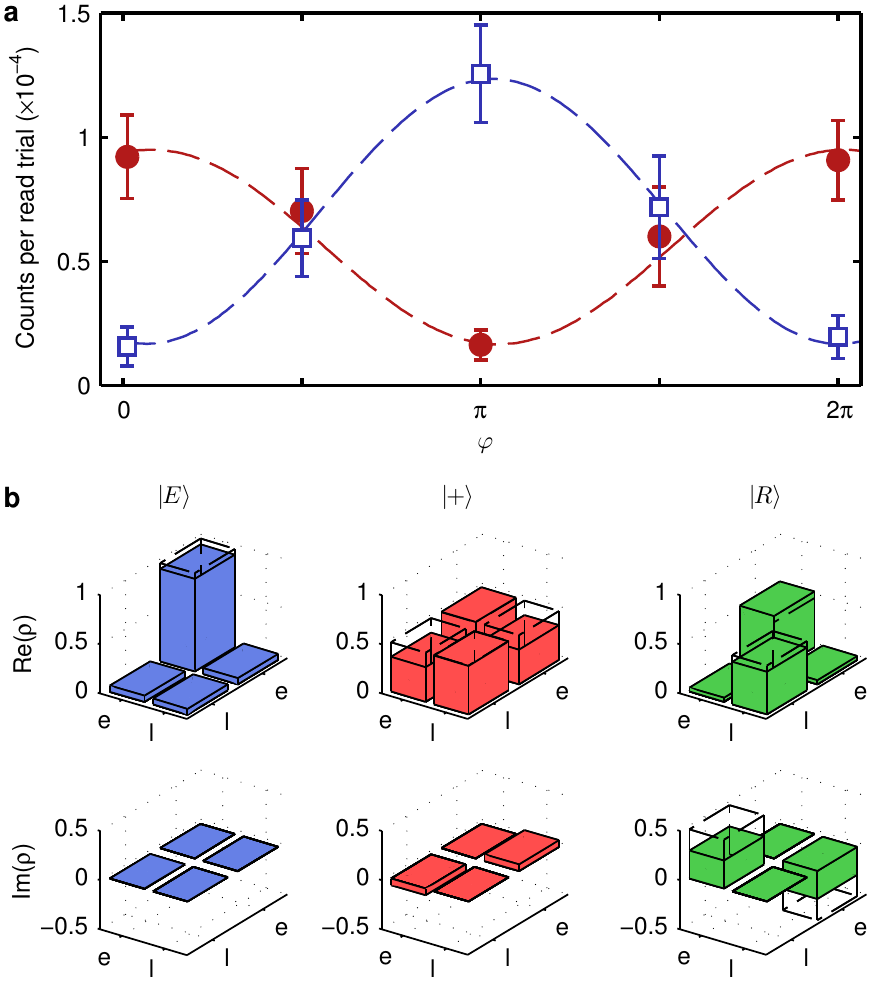}
\caption{\textbf{Single photon qubit transfer.} (a) Interference fringes from a single time-bin read photon $|\Psi_\mathrm{eq}\rangle=\frac{1}{\sqrt{2}}(|e\rangle+e^{i\varphi}|l\rangle)$ generated at site \textit{A} ($p_\mathrm{e}\approx 5\%$ corresponding to $\alpha=0.26$) if the second AFC is prepared with zero detuning (red dots, $V=70\pm6\%$) or shifted by $200\,\mathrm{kHz}$ (blue open squares, $V=76\pm3\%$). Error bars correspond to $\pm1\,\mathrm{s.d.}$ of the photon counting statistics. (b) Real and imaginary parts of the reconstructed density matrices measured after the crystal at site \textit{B} if the qubits $|E\rangle$, $|+\rangle$, $|R\rangle$ are generated at site \textit{A}, converted in the QFCDs and stored at \textit{B}. Open boxes indicate the target state.}
\label{Figure4}
\end{figure}

Finally, to demonstrate qubit transfer between the cold atomic cloud and the crystal via telecom photons, we decreased $p_\mathrm{e}$ to $5\%$ generating true single time-bin read photons at site \textit{A} with an anti-bunching parameter of $\alpha=0.26\pm0.02$ (cf. Fig.~\ref{Figure2}(a)). Following the same approach as above, we show in Fig.~\ref{Figure4}(a) that with converted and stored single time-bin photons, we obtain interference between overlapping bins with visibilities around $70\%$. Moreover, we show that, by changing the central frequency of the second AFC by $200\,\mathrm{kHz}$, the phase of the interference fringe can be shifted by $180^\circ$ verifying the intrinsic phase analyzing capabilities of the AFC (see Methods). This capability permits the measurement of time-bin qubits in different bases and hence a full analysis of the stored qubits by quantum state tomography. Figure~\ref{Figure4}(b) shows the reconstructed density matrices $\rho$ of the retrieved states after the crystal at site \textit{B} when three orthogonal time-bin qubits ($|E\rangle$=$|e\rangle$, $|+\rangle=1/\sqrt{2}(|e\rangle+|l\rangle)$, $|R\rangle=1/\sqrt{2}(|e\rangle+i|l\rangle)$) are generated in the cold atomic QM at site \textit{A}, afterwards converted in the QFCDs, and stored at site \textit{B}. The state reconstruction is based on maximum likelihood estimation. The qubit fidelity conditioned on a successful detection of the photon after the retrieval from the crystal (conditional fidelity) is calculated as  $\mathcal{F}^c_{|\psi\rangle}=\langle\psi|\rho|\psi\rangle$ with $|\psi\rangle$ denoting the target state. From Fig.~\ref{Figure4}(b) it is evident that we obtain a high overlap between the reconstructed qubits and the target states with conditional fidelities of $\mathcal{F}^c_{|+\rangle}=85.4\pm6.6\%$, $\mathcal{F}^c_{|R\rangle}=78.2\pm6.9\%$, and $\mathcal{F}^c_{|E\rangle}=93.8\pm2.8\%$, where the errors were estimated via Monte Carlo simulations taking into account the uncertainty of the photon counting statistics. Despite the low total efficiency of the state transfer we demonstrate an average conditional fidelity of $\mathcal{F}^c=85.8\pm3.3\%$ for the generated and transferred qubit which is consistent with results inferred at higher $p_\mathrm{e}$ and surpasses the classical threshold of $66.7\%$ by more than 5 standard deviations. 
Overall, the conditional fidelities are limited by the SNR of the retrieved photons, thus by the efficiencies of all the involved processes. However, for the equatorial states, the main limitation is the finite linewidth of the combined laser system, which we estimate to be around  600 kHz (see Methods). This large sensitivity to frequency fluctuations is due to the large separation between the two time-bins ($500\,\mathrm{ns}$) imposed by the AFC bandwidth.

For potential applications in hybrid quantum networks, the transfer efficiency (currently $10^{-3}$) should be greatly increased. Note that the largest part of the inefficiency is due to technical optical loss in the various elements ($\eta_{loss}=0.04$). This could be significantly improved, using e.g. fiber pigtailed waveguide converters. The combined quantum memory efficiency ($\eta_{QM}=0.09$) could also be largely increased with state of the art techniques \cite{Hedges2010, Yang2016}. Increased efficiencies would also enable spin-wave storage in the crystal, leading to on-demand read-out and longer storage times \cite{Seri2017}. While all efficiencies could be in principle pushed towards unity, an interesting direction to alleviate optical losses would be to implement a non-destructive detection of the time-bin qubit with the AFC, as recently proposed in ref. \cite{Sinclair2016}.

Our work represents a demonstration of quantum communication between heterogeneous quantum nodes and opens prospects for combining quantum nodes with different capabilities. Moreover, it gives a perspective on how the distance between the nodes can be extended by back- and forth-conversion of photonic qubits into the telecom C-band. Our technique could also be extended to connect other physical platforms, e.g. single ions or NV centers. Our results hold promise for the realization of large scale hybrid quantum networks.



\section{Acknowledgements}

Research at ICFO is supported by the ERC starting grant QuLIMA, by the Spanish Ministry of Economy and Competitiveness (MINECO) and the Fondo Europeo de Desarrollo Regional (FEDER) through grant FIS2015-69535-R, by MINECO Severo Ochoa through grant SEV-2015-0522, by AGAUR via 2014 SGR 1554,  by Fundaci\'{o}  Privada Cellex and by the CERCA programme of the generalitat de Catalunya. P.F. acknowledges the International PhD-fellowship program "la Caixa"-Severo Ochoa @ ICFO. G.H. acknowledges support by the ICFOnest international postdoctoral fellowship program.\\

\section{Author contributions}
 N.M. built and operated the QFC setups, P.F. built and operated the atomic QM setup, both under supervision of G.H. The solid state QM setup was built and operated by K.K. under supervision of M.M. The experiment was conducted by N.M., P.F., K.K., and G.H., who also jointly analyzed the data. G.H., N.M and H.d.R. wrote the paper, with inputs from all co-authors. H.d.R conceived the experiment and supervised the project.

\newpage
\pagebreak 
\newpage

\newpage


\section{Methods}

{\small

In the following, we give a detailed description of the experimental procedures and the experimental setup as shown in Figure 5. We also present additional measurements and discuss limitations. 

\begin{figure*}[htbp]
\includegraphics[width=1.0\textwidth]{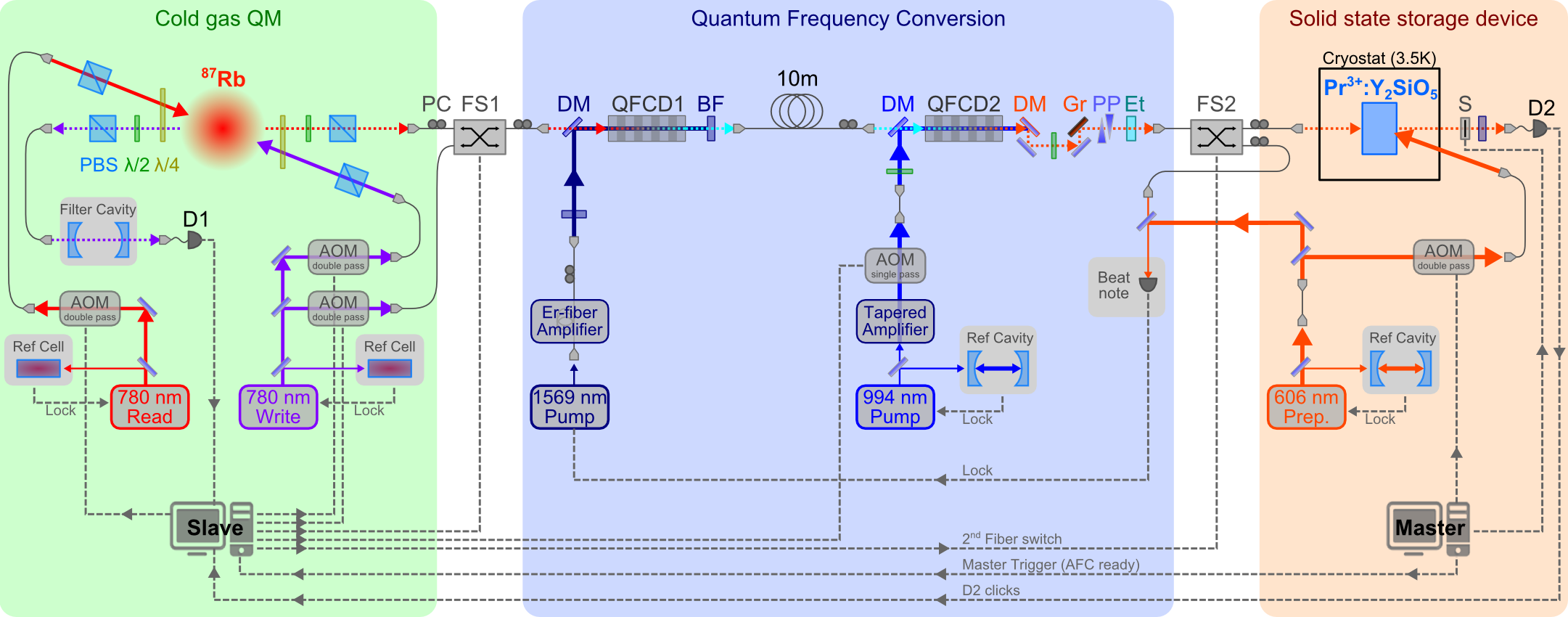}
\caption{\textbf{Experimental setup.} On the left, a cold $^{87}\mathrm{Rb}$ QM is operated to generate non-classically correlated photon pairs. In the center, the single read photons are frequency converted from $780\,\mathrm{nm}$ to $1550\,\mathrm{nm}$ in QFCD1 and afterwards to $606\,\mathrm{nm}$ in QFCD2. On the right, the converted read photons are stored and analysed in a crystal before being retrieved and detected. Abbreviations: PBS polarizing beam splitter; $\lambda/2$ ($\lambda/4$) half (quarter) wave plate; AOM acousto optic modulator; PC polarization controller; FS fiber switch; DM dichroic mirror; BF band pass filter; Gr diffraction grating; PP anamorphic prism pair; Et etalon; S mechanical shutter.}
\label{FigureSM1}
\end{figure*}

It is convenient to divide the experiment into three sections --- I. the cold atomic gas Quantum Memory (QM), II. the solid state storage device, III. the Quantum Frequency Conversion (QFC) interface. We first discuss these three sections separately, before addressing additional measurements in section~IV.


\section{I. Cold Atomic Gas QM}
\label{sec:ColdGasQM}

\subsection{Setup}
\label{subsec:ColdGasQM-Setup}
The cold atomic QM at site~\textit{A} consists of a cloud of $^{87}\mathrm{Rb}$ atoms, kept in a UHV chamber and cooled via magneto optical trapping. The cooling and repumping beams (not shown in Figure 5) are derived from the write and read diode lasers which are locked via doppler free absorption spectroscopy to Rubidium reference cells to be resonant with the $D_2$ line of $^{87}\mathrm{Rb}$ at $780\,\mathrm{nm}$. After passing through acousto optic modulators (AOMs) in double pass configuration, the cooling beam is $20\,\mathrm{MHz}$ red-detuned to the ${|F=2\rangle\leftrightarrow|F'=3\rangle}$ transition, and the repumping beam is resonant to the ${|F=1\rangle\leftrightarrow|F'=2\rangle}$ transition. They are combined with a magnetic gradient of $20\,\mathrm{G/cm}$ to load $N\approx 10^8$ Rubidium atoms into the MOT. After a $2\,\mathrm{ms}$ long optical molasses phase, the temperature of the atoms is about $T\approx100\,\mu\mathrm{K}$. Next, all population is prepared in the $|g_{A}\rangle=|5S_{1/2},F=2,m_{F}=2\rangle$ Zeeman sublevel by applying the repumping light and the $\sigma^+$ polarized optical pumping light on the ${|F=2\rangle\rightarrow|F'=2\rangle}$ transition (not shown in Figure 5).

To generate the spin-wave inside the atomic cloud, we send write pulses (derived from the write laser and a subsequent AOM) which are $40\,\mathrm{MHz}$ red detuned from the $|g_{A}\rangle \rightarrow |e_{A}\rangle=|5P_{3/2},F=2,m_{F}=1\rangle$ transition and exhibit a duration of $20\,\mathrm{ns}$ (full width half maximum FWHM). The write pulses pass a polarization beam splitter (PBS) and a quarter wave plate to set their polarization to $\sigma^-$ in the frame of the atoms. The quantization axis is set by a bias magnetic field of $B=110\,\mathrm{mG}$ along the write/read photon direction. The write pulses generate Raman-scattered write photons, which are emitted on transition $|e_{A}\rangle \rightarrow |s_{A}\rangle=|5S_{1/2},F=1,m_{F}=0\rangle$. A small fraction of the isotropically emitted write photons is collected at an angle of $3.4^\circ$ with respect to the write/read pulse axis. The write photons pass a combination of quarter wave plate, half wave plate, and PBS to couple just the ones with the correct $\sigma^+$ polarization into a polarization maintaining (PM) optical fiber. Afterwards the write photons are spectrally filtered by a monolithic Fabry-Perot cavity with approx. $50\,\mathrm{MHz}$ linewidth and $24\%$ total transmission (including subsequent fiber couling) before finally being detected by SPD D1 with $41\%$ efficiency and a dark count rate of $130\,\mathrm{Hz}$.

To read-out the atomic spin-wave, we send after a storage time of $\tau_A=1.6\,\mu\mathrm{s}$ \textit{read} pulses (derived from the read laser and a subsequent AOM) which are resonant with the $|s_{A}\rangle \rightarrow |e_{A}\rangle$ transition. They propagate in the same optical mode but opposite direction than the write pulses and their polarization is set to $\sigma^+$ by a PBS and a quarter waveplate. The intensity and temporal wave shape of the read pulses are tailored to efficiently generate read photons with tunable waveform \cite{Farrera2016a}.
The read photons are emitted on the $|e_{A}\rangle \rightarrow |g_{A}\rangle$ transition and leave the atomic cloud in the opposite direction than the write photons due to the phase matching condition $\vec{k}_{r}=\vec{k}_{R}+\vec{k}_{W}-\vec{k}_{w}$, where $\vec{k}_{r}$, $\vec{k}_{w}$ ($\vec{k}_{R}$, $\vec{k}_{W}$) represent the wave vectors of the photonic (pulse) modes. The polarization of the read photons is $\sigma^-$ when leaving the atomic cloud and is subsequently changed to linear by a quarter wave plate. The read photons are filtered by a combination of half wave plate and PBS before they are coupled with an efficiency of approximately $60\%$ into a PM fiber. The fiber is connected to a micro-electro-mechanical single-mode fiber-optic switch (FS1) which directs the read photons or classical lock light to the QFC interface (see section~III).

\begin{figure}
\includegraphics[width=1.0\columnwidth]{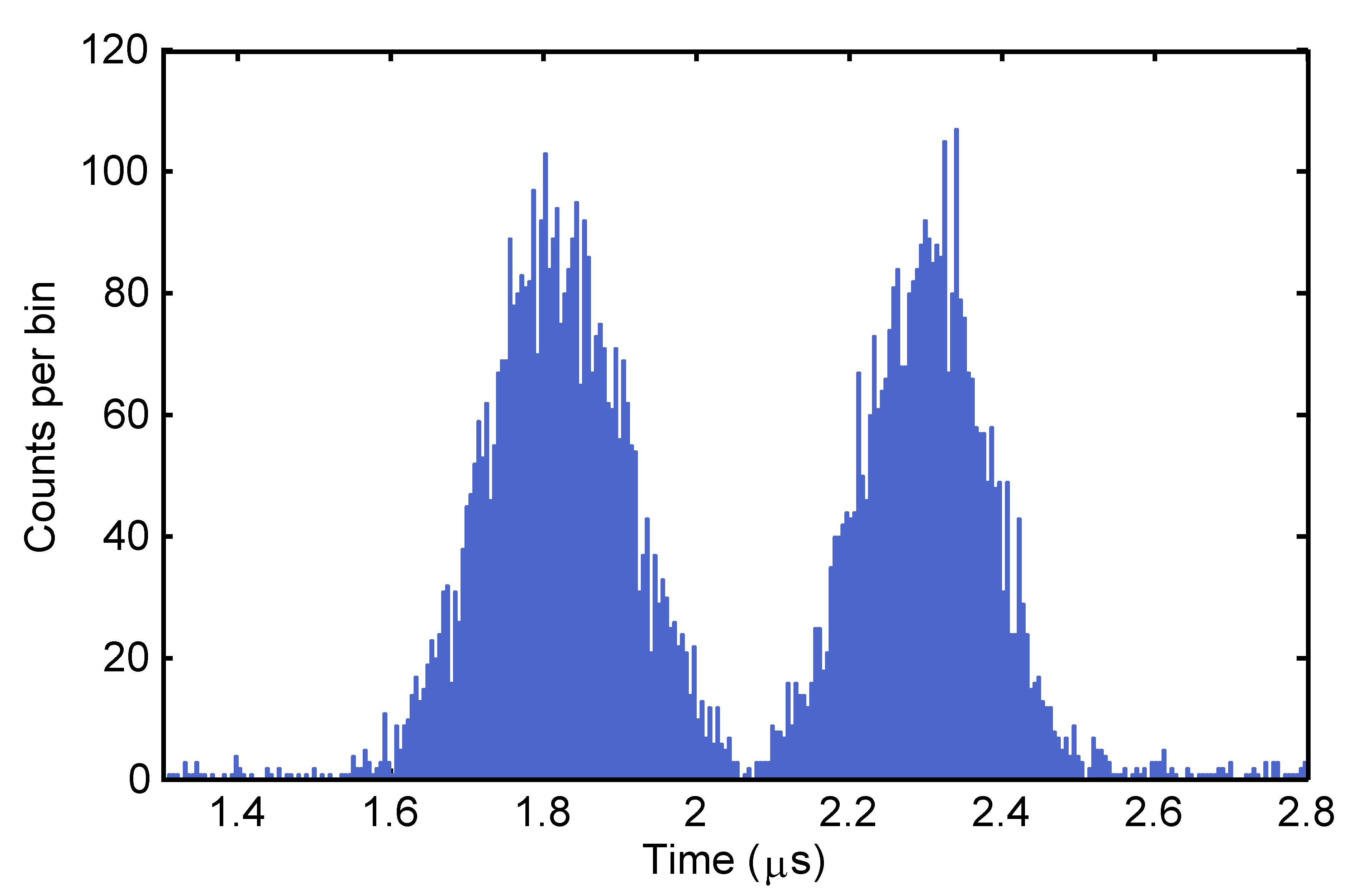}
\caption{\textbf{Time-bin read photon.} Conditional histogram of a time-bin read photon, taken at $p_\mathrm{e}=5\%$ after the MOT at site~\textit{A}.}
\label{TimeBinPhoton}
\end{figure}

To generate sub-natural linewidth single photons in the cold atomic QM which exhibit a temporally delocalized wave shape suitable for encoding photonic time-bin qubits, we follow the approach described in \cite{Farrera2016a}. If a heralding write photon is detected at D1, we map the spin-wave in the Rb QM onto a photonic time-bin qubit depending on the temporal shape of the read pulse. Instead of sending a simple Gaussian shaped read-out pulse to the cold atomic ensemble, we apply an appropriately imbalanced doubly-peaked read-out pulse. The first (early) peak reads out the stored spin-wave with half the retrieval efficiency $\eta_\mathrm{ret}^{A}/2$, and the second (late) peak with full retrieval efficiency $\eta_\mathrm{ret}^{A}$. This creates the desired time-bin read photon with equal photon detection probabilities in both time-bins ($c_1=c_2=\frac{1}{\sqrt{2}}$). By controlling the phase $\varphi$ between both read-out peaks we can thus create a time-bin photon representing an equatorial qubit state $|\Psi_\mathrm{eq}\rangle=\frac{1}{\sqrt{2}}(|e\rangle+e^{i\varphi}|l\rangle)$.

Moreover, by changing the duration of the read-out peaks, we are able to precisely tune the duration of the whole photon or both time-bins individually. We are thereby able to create single photons which exhibit sub-natural linewidths in the range of $2\,\mathrm{MHz}$, matching the spectral requirements of the AFC memory. This also allows us to generate two time-bins with identical shape as needed for high visibility interference for the coherence preservation and qubit analysis experiments.

Figure 6 shows an example histogram of a time-bin read photon generated at $p_\mathrm{e}=5\%$ in the cold atomic QM detected right after the MOT at site~\textit{A}. Characterization of that photon via a Hanbury Brown Twiss setup after the MOT yields a heralded autocorrelation function of $g^{(2)}_{r_A,r_A|w}=0.26\pm0.02$ confirming the single photon nature of that time-bin read photon.

\section{II. Solid state storage device}
\label{sec:SolidStateQM}
\subsection{Setup}
\label{subsec:SolidStateQM-Setup}
\label{AFC}
The solid state storage device at site~\textit{B} is a bulk $\mathrm{Pr}^{3+}\mathrm{:Y}_2\mathrm{SiO}_5$ crystal (Scientific Materials) cooled down to $3.5\,\mathrm{K}$ in a cryostat (Cryostation, Montana Instruments). With a Pr$^{3+}$ ion concentration of $0.05\%$ and a length of $5\,\mathrm{mm}$, the crystal features a total optical depth of about $10$ at the frequency of the $^{3}\mathrm{H}_4(0) \leftrightarrow{}^{1}\mathrm{D}_{2}(0)$ transition (see Fig.~1 in the main text). 
The laser used to address this transition at $606\,\mathrm{nm}$ is a Toptica DL SHG pro, stabilized via the Pound-Drever-Hall technique to a home-made Fabry-Perot cavity in vacuum. 
From that laser we derive the reference beam for the QFC frequency locking (see section~III) and the beam necessary to prepare and operate the solid state memory. The latter beam is modulated in amplitude and frequency by means of a double pass AOM and is guided to the cryostat through a PM single-mode fiber. It finally arrives at the crystal with a waist of $300\,\mathrm{\mu m}$ and an angle of $\sim 4\,\mathrm{^{\circ}}$ with respect to the direction of the converted single photons. These, instead, have a waist at the crystal of $40\,\mathrm{\mu m}$. 

The chosen technique to store the converted single photons is the atomic frequency comb (AFC) protocol \cite{Afzelius2009}. It relies on the preparation of a periodic absorptive structures within the inhomogeneously broadened absorption profile of Pr$^{3+}$ which is able to maintain the coherence of the absorbed photons. As a matter of fact, a photon absorbed by an AFC with periodicity $\Delta$ is mapped into a coherent superposition of atomic excitations that, after a time $\tau_B = 1/\Delta$, experiences a rephasing and gives rise to a collective re-emission in forward direction.

To reshape the absorption profile of the $\mathrm{Pr}^{3+}\mathrm{:Y}_2\mathrm{SiO}_5$ crystal into an AFC, we follow a procedure similar to that described in \cite{Gundogan2015}. We first sweep the laser by $12\,\mathrm{MHz}$ with a power of $24\,\mathrm{mW}$. This empties the $|\pm1/2\rangle_g$ and $|\pm3/2\rangle_g$ hyperfine states of a certain class of atoms and creates a $18\,\mathrm{MHz}$-wide transparency window. We then send $4\,\mathrm{MHz}$ broad \textit{burn-back} pulses resonant with the transition $|\pm5/2\rangle_g \leftrightarrow |\pm5/2\rangle_e$ to repump back atoms in the state $|\pm1/2\rangle_g$. Subsequent $5\,\mathrm{MHz}$-broad pulses resonant to the $|\pm3/2\rangle_g \leftrightarrow |\pm3/2\rangle_e$ transition clean the $|\pm3/2\rangle_g $ state which was also repopulated by the \textit{burn-back} pulses. Moreover, the cleaning pulses have the secondary effect of removing from the frequency range of interest any absorption feature associated to other atomic classes. At this stage, we have a $4\,\mathrm{MHz}$-wide single class absorption feature resonant to the transition $|\pm1/2\rangle_g \leftrightarrow |\pm3/2\rangle_e$, where we finally prepare the atomic frequency comb.
This is done by burning spectral holes with repetitions of low power pulses whose frequency is changed by a fixed amount $\Delta$. For the storage of time-bin qubits we repeat the latter operation twice with two different periodicities, $\Delta_1$ and $\Delta_2$, such that the AFC gives access to two different storage times, $\tau_{B1}$ and $\tau_{B2}$. Figure 7 shows the combs structures and their corresponding echos for the different storage times used in the experiment. 

\begin{figure}
\includegraphics[width=1.0\columnwidth]{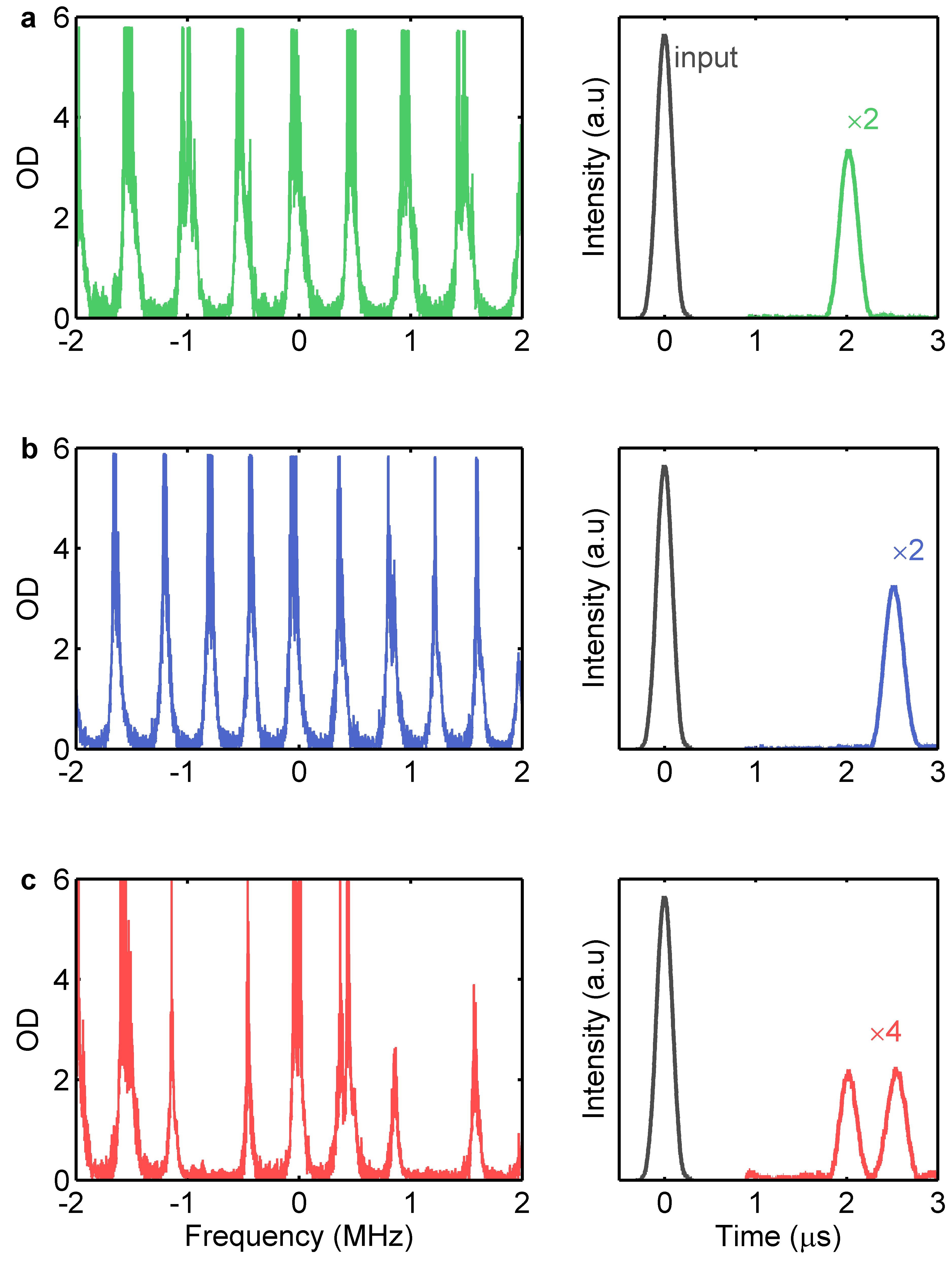}
\caption{\textbf{AFC storage characterization.} Different absorption spectra of atomic frequency combs with different periodicities $\Delta$ are shown on the left side. $200\,\mathrm{ns}$ FHWM input pulses derived from the $606\,\mathrm{nm}$ preparation laser are sent to the different AFC structures at 0 $\mu s$ and their corresponding echos are shown on the right side. (a) $\Delta=500\,\mathrm{kHz}$ (b) $\Delta=400\,\mathrm{kHz}$ (c) double periodicity with $\Delta_1=400\,\mathrm{kHz}$ and $\Delta_2=500\,\mathrm{kHz}$ leading to a double echo at $2\,\mu\mathrm{s}$ and $2.5\,\mu\mathrm{s}$}
\label{AFCs}
\end{figure}

To protect the single photon detector from the leakage of the preparation beam, we employ a mechanical shutter after the cryostat which remains closed during the whole AFC preparation. Also, the retrieved photons pass through a band-pass filter (centered at $600\,\mathrm{nm}$, linewidth $10 \,\mathrm{nm}$, Semrock) before reaching the detection stage, implemented with the SPD D2 (PicoQuant, $45\,\%$ detection efficiency and $15\,\mathrm{Hz}$ dark-count rate).

\subsection{Qubit Analysis}
To analyze the phase of the qubit, we use the AFC as an interferometric device \cite{DeRiedmatten2008}. By preparing two AFCs with different periodicities $\Delta_1$ and $\Delta_2$, leading to storage times of $\tau_{B_1}=2\,\mu\mathrm{s}$ and $\tau_{B_2}=2.5\,\mu\mathrm{s}$ (corresponding to the short and long path of the interferometer), the early echo of the late time-bin and late echo of the early time-bin overlap and interfere. To control the phase of one arm of the interferometer, we exploit the fact that the emitted echo from an AFC acquires a phase shift of $e^{i2\pi\delta/\Delta}$, where $\delta$ is the frequency detuning between the center of the AFC and the input photon, and $\Delta$ the periodicity of the comb \cite{Afzelius2009}. Hence, shifting $\delta$ for one of the two AFCs, allows full control of the interferometer and permits us to set the measurement basis.

\subsection{Density matrices}
Here we provide the numerical result of the quantum state tomography \cite{James2001} in the form of density matrices of the three investigated states estimated via maximum likelihood estimation. Based on a Monte Carlo simulation we obtain a $5\sigma$ violation of the $66.7\%$ classical threshold \cite{Massar1995}.
\begin{equation}
\rho_R = \begin{bmatrix}
	0.567 & 0.040 - i\,0.284\\
	0.040 + i\,0.284  & 0.434
	\end{bmatrix}
\end{equation}
\begin{equation}
\rho_+ = \begin{bmatrix}
0.505 &0.352 + i\,0.067\\
0.352 - i\,0.067  & 0.495
\end{bmatrix}
\end{equation}
\begin{equation}
\rho_E = \begin{bmatrix}
0.941 & 0.063 - i\,0.001\\
0.063 + i\,0.001  & 0.059
\end{bmatrix}
\end{equation}

\subsection{AFC vs. photon detuning}
\label{subsec:SolidStateQM-AFCdetuning}
\begin{figure}
\includegraphics[width=1.0\columnwidth]{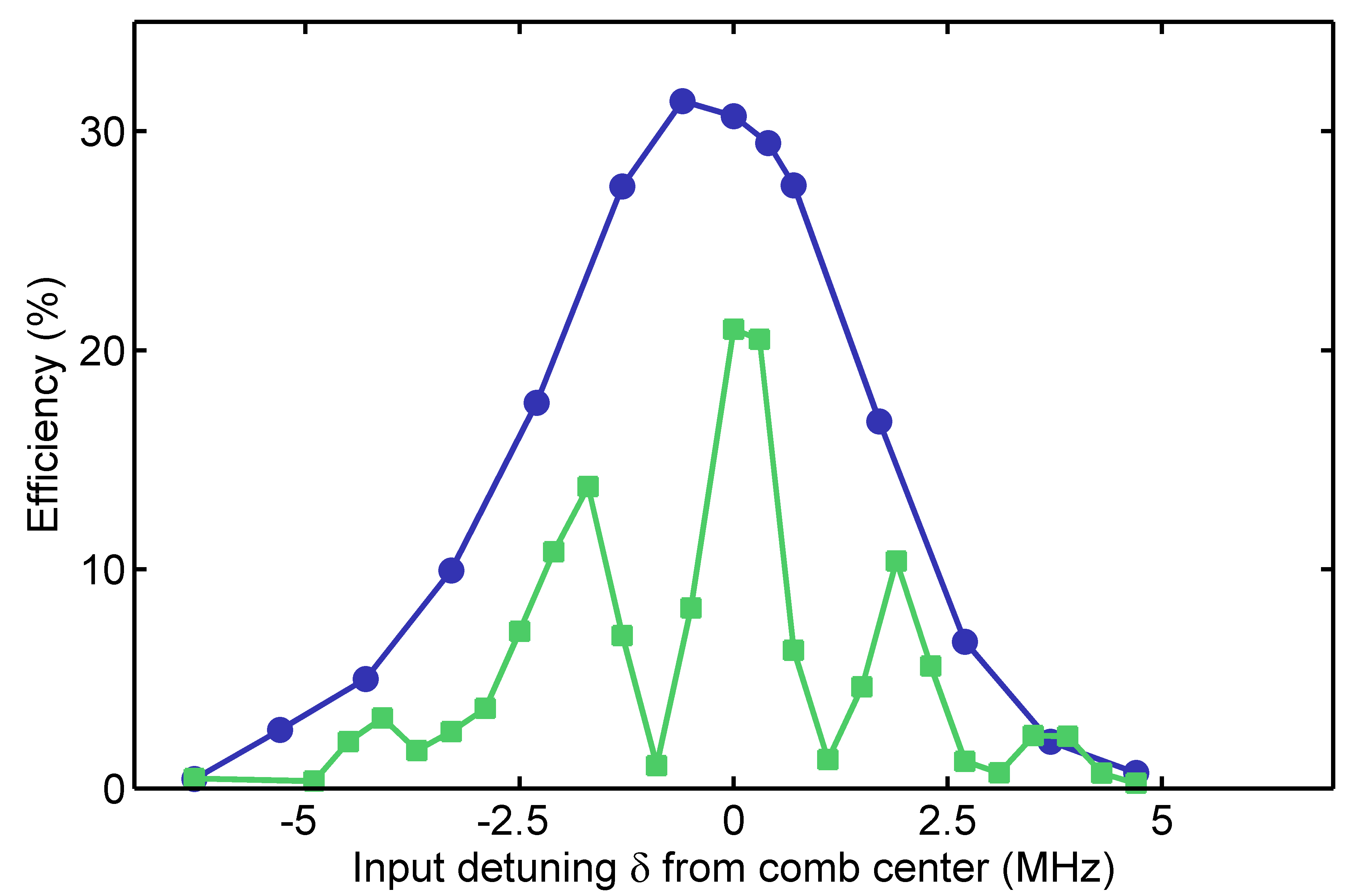}
\caption{\textbf{AFC efficiency and interference versus photon detuning.} Relative storage efficiency of classical light (derived from the $606\,\mathrm{nm}$ preparation laser) versus its frequency detuning with respect to the center of the prepared AFC. (blue dots) The input is a single Gaussian shaped pulse stored in a single AFC. (green squares) The input is a doubly-peaked pulse (mimicking a time-bin input photon) stored on two superimposed AFCs.}
\label{Detuning}
\end{figure}

The AFC storage efficiency as a function of the frequency detuning $\delta$ of the input light from the center of the comb is investigated in Figure 8. The blue dots show the storage efficiency of a $606\,\mathrm{nm}$ single classical pulses with $200\,\mathrm{ns}$ FWHM, on a $4\,\mathrm{MHz}$ wide comb for a storage time of $\tau_{B}=2.5\,\mu\mathrm{s}$. As expected, the storage efficiency drops significantly when the input light is not anymore resonant with the prepared AFC.

Next, double Gaussian pulses separated by $500\,\mathrm{ns}$, mimicking a time-bin qubit with zero relative phase, are stored on two superimposed combs with $2\,\mu\mathrm{s}$ and $2.5\,\mu\mathrm{s}$ storage times, such that the late echo of the early bin and the early echo of the late bin overlap in time and interfere. The green squares show the area of the interfering peak relative to the total input pulse, as a function of the input light frequency detuning $\delta$. It shows an oscillating behaviour corresponding to constructive and destructive interferences caused by the frequency detuning of the input light pulses to the center of the combs, inducing a relative phase shift of $e^{i2\pi\left(\frac{\delta}{\Delta_1}-\frac{\delta}{\Delta_2}\right)} = e^{i2\pi\delta(\tau_1-\tau_2)}$ between the two interfering echoes. As expected, a $2\pi$ phase shift is observed for an input frequency detuning $\delta$ of $2\, \mathrm{MHz}$ corresponding to the $500\, \mathrm{ns}$ separation between the early and late bin.

This measurement highlights the importance of laser stabilization in order to achieve high storage efficiency and strong coherence conservation (see section~III).

\section{III. The Interface}
\label{sec:Interface}

\subsection{Quantum Frequency Conversion Setup}
\label{subsec:Interface-Setup}
The interface between the cold atomic ensemble -- emitting read photons at $780\,\mathrm{nm}$ wavelength -- and the crystal {\textendash} storing photons at $606\,\mathrm{nm}$ {\textendash} is based on two quantum frequency conversion devices (QFCDs) \cite{Kumar1990}. The first QFCD shifts the frequency of $780\,\mathrm{nm}$ light to the telecom C-band by difference frequency generation (DFG). It is based on a proton exchange (PE) waveguide inside a periodically poled lithium niobate (PPLN) chip, where $780\,\mathrm{nm}$ light is coupled together with $1569\,\mathrm{nm}$ pump light derived from an Erbium doped fiber amplifier fed by an external cavity diode laser. Both light fields are combined on a dichroic mirror and coupled into the waveguide by an aspheric lens with efficiencies of $\eta^\mathrm{780\,nm}_\mathrm{cpl}=44\%$ and $\eta^\mathrm{1569\,nm}_\mathrm{cpl}=36\%$. Inside the temperature stabilized waveguide, $290\,\mathrm{mW}$ of pump light convert $780\,\mathrm{nm}$ light to $1552\,\mathrm{nm}$ via DFG, with an internal conversion efficiency of $\eta_\mathrm{int}^\mathrm{QFCD1}=56\%$. After collimating the waveguide output, a combination of two band pass filters (BF), each with a transmission bandwidth of $7\,\mathrm{nm}$ around $1552\,\mathrm{nm}$ and a maximum optical depth of $OD \approx 12$ at $1569\,\mathrm{nm}$, separates the converted light from the pump light. The combined transmission of the BF and the subsequent coupling into a single mode telecom fiber is $68\%$ giving a total device efficiency of QFCD1 of $\eta_\mathrm{dev}^\mathrm{QFCD1}=17\%$. 

After passing FS1, the converted light is then sent to a second laboratory via a $10$ meter telecom fiber, where the second QFCD shifts the frequency of $1552\,\mathrm{nm}$ light to $606\,\mathrm{nm}$ by Sum Frequency Generation (SFG). QFCD2 is based on a temperature stabilized Ridge-waveguide made of PPLN. The waveguide is pumped by $994\,\mathrm{nm}$ radiation, derived from an external cavity diode laser locked via the Pound Drever Hall technique to an external reference cavity with $690\,\mathrm{kHz}$ linewidth and a free spectral range (FSR) of $1\,\mathrm{GHz}$ in order to reduce its linewidth. The pump light is amplified by a tapered amplifier, sent through a gating AOM and an optical fiber to clean the spatial mode. After its polarization is adjusted by a half wave plate, the pump light is combined with the $1552\,\mathrm{nm}$ telecom light on a dichroic mirror and coupled into the waveguide by an aspheric lens with efficiencies of $\eta^\mathrm{994\,nm}_\mathrm{cpl}=62\%$ and $\eta^\mathrm{1552\,nm}_\mathrm{cpl}=51\%$, respectively. Inside the waveguide $450\,\mathrm{mW}$ of pump field convert the $1552\,\mathrm{nm}$ light to $606\,\mathrm{nm}$ by means of SFG with an internal conversion efficiency of $\eta_\mathrm{int}^\mathrm{QFCD2}=60\%$. After collimating the waveguide output, several optical elements are used to filter the converted light at $606\,\mathrm{nm}$ from the pump radiation, the Raman noise from the first QFCD, and the pump SPDC noise generated and converted in the second QFCD. First, a dichroic mirror (transmission $T_\mathrm{DM}^{1552\,\mathrm{nm}}=94\%$) is used, before the light is sent to a diffraction grating (diffraction efficiency $\eta_\mathrm{Gr}^{606\,\mathrm{nm}}=75\%$) and an etalon ($T_\mathrm{Et}^{606\,\mathrm{nm}}=95\%$, $\mathrm{FSR}=60\,\mathrm{GHz}$, finesse $\mathcal{F}=6$). In between, a half wave plate and an anamorphic prism pair adjust the polarization for the grating and the spatial mode of the $606\,\mathrm{nm}$ light to efficiently couple it into the second fiber switch (FS2) which directs the converted light to the solid state storage device. The combined transmission of the filtering elements and the subsequent coupling into the single mode fiber of FS2 is $48\%$ giving a total device efficiency of QFCD2 of $\eta_\mathrm{dev}^\mathrm{QFCD2}=15\%$. 

The probability for a $780\,\mathrm{nm}$ photon entering the first fiber switch FS1 and to exit the second fiber switch FS2, at $606\,\mathrm{nm}$ is about $1.2\%$. This includes all possible losses: conversion efficiencies, waveguide coupling, fiber coupling, optical filtering and transmissions (see Table 1).

\begin{table}[htb]
	
	\caption{\textbf{System losses.} Detailed optical transmissions and efficiencies of the experiment.}
	
	
	\centering
	
	\begin{tabular}{ K{1.3cm} | K{3.5cm}  K{1cm} | K{2cm}}
						&  								& $\mathrm{T,}\,\eta$\\
		\hline	
		Cold gas		& read retrieval (in fiber)  	& $30\%$&\\ 	
		\hline	
		FS1				& transmission			  		& $72\%$&\\ 
		\hline
						& waveguide coupling  			& $44\%$&\\ 
		QFCD1			& conversion			  		& $56\%$& $\eta_{\mathrm{device}}=\mathrm{17}\%$\\ 
						& filtering \& fiber coupling  	& $68\%$&\\ 
		\hline
						& waveguide coupling  			& $51\%$&\\ 
		QFCD2			& conversion  					& $60\%$& $\eta_{\mathrm{device}}=\mathrm{15}\%$\\ 
						& filtering  					& $75\%$&\\ 
						& fiber coupling 		  		& $64\%$&\\ 
		\hline		
		FS2				& transmission  				& $70\%$&\\ 
		\hline
						& AFC storage  					& $29\%$&\\ 
		Crystal			& optical transmission  		& $52\%$&\\ 
						& detection  					& $45\%$&\\ 
		\hline			
	\end{tabular}
	
	
	\label{Table2}
	
\end{table}

\subsection{Frequency Locking Scheme}
\label{subsec:Interface-LockingScheme}
Active stabilization of the involved laser frequencies is necessary to ensure that the converted read photons emitted by the cold atomic QM are resonant to the AFC structure prepared in the crystal.
Herefore, the conversion interface is used in two different configurations: a ‘QFC’ mode where the read photons are converted and sent to the AFC memory, and a ‘LOCK’ mode where 780 nm continuous wave (CW) light, derived from the write laser, is converted and used to stabilize the frequency of the converted read photons. Two single mode fiber switches (FS1 \& FS2) placed before and after the interface are used to swap between the two modes. 

The first FS placed before the interface has two inputs and one output and the second one after the interface has one input and two outputs. In the QFC mode, the first FS couples the $780\,\mathrm{nm}$ read photons to the frequency converters and the second switch directs the converted $606\,\mathrm{nm}$ photons to the solid state storage device. In the LOCK mode, FS1 couples $2\,\mathrm{mW}$ of $780\,\mathrm{nm}$ CW light to the interface and FS2 sends the converted light to the lock system. An optical beat note between the converted ‘LOCK’ light and the reference $606\,\mathrm{nm}$ laser (used to prepare the AFC structure in the crystal) is measured using a photodiode. The beat note is stabilized at $104\,\mathrm{MHz}$ using a frequency comparator (based on a phase locked loop referenced to an internal clock) which feeds back an error signal to the $1569\,\mathrm{nm}$ pump laser. Any drift of the involved lasers, inducing a frequency shift of the converted photons is then compensated by acting on the current of the $1569\,\mathrm{nm}$ pump laser, thus ensuring that the later converted $606\,\mathrm{nm}$ read photons are resonant to the AFC structure in the crystal. 

\subsection{Experimental Time Sequence}
\label{subsec:Interface-TimeSequence}
\begin{figure}
\includegraphics[width=1.0\columnwidth]{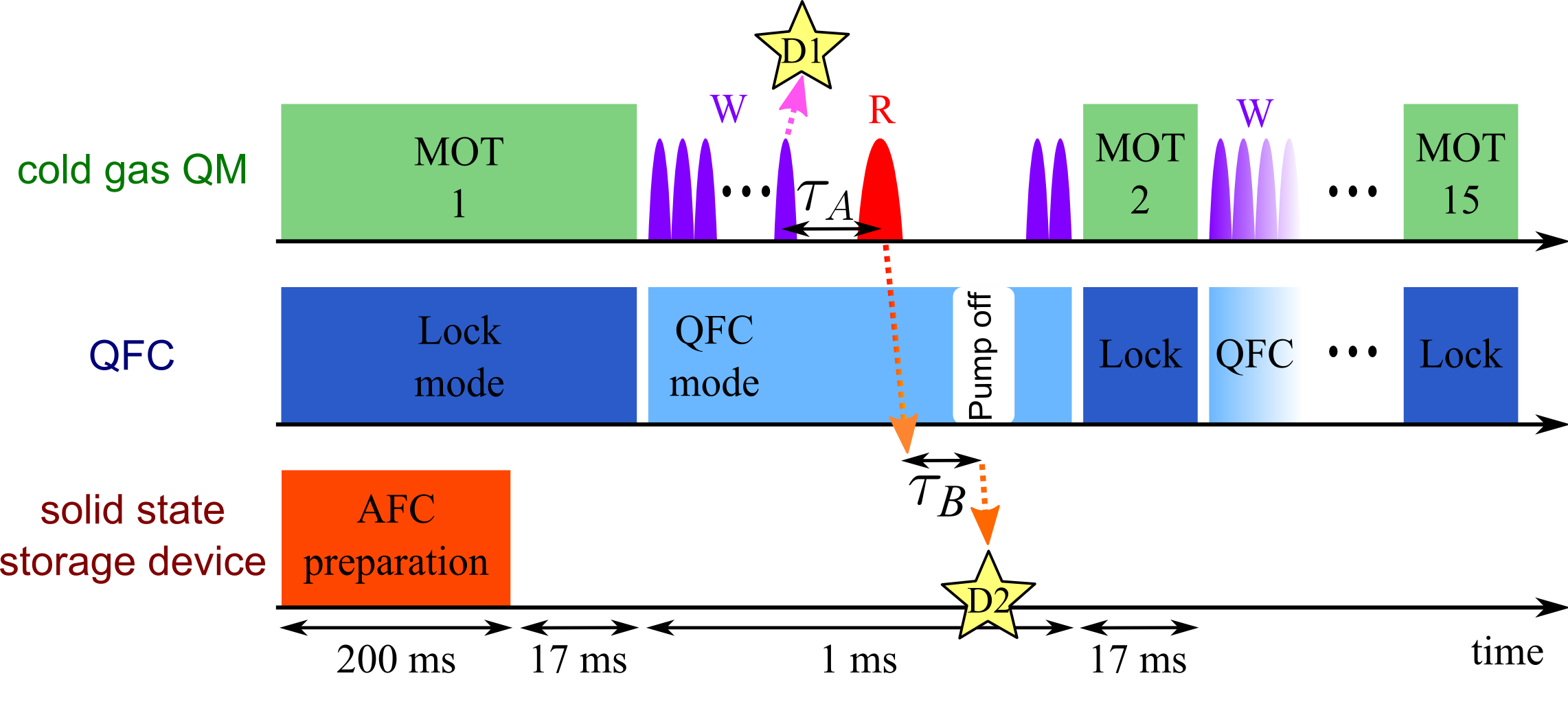}
\caption{\textbf{Experimental time sequence.} First the AFC in the crystal is prepared (bottom row) before the main experiment involving the cold atomic QM (top row) and the conversion interface (center row) starts. Eventual detections of write photons at D1 and converted, stored and restored read photons at D2 are indicated by stars.}
\label{sequence}
\end{figure}
The time sequence for the experiments presented in the main text is shown in Figure 9. Synchronized on the cryostat cycle of $1\,\mathrm{Hz}$, it starts by preparing the AFC in the crystal for up to $200\,\mathrm{ms}$. Once completed, the Master Computer sends a trigger to the Slave Computer that controls the rest of the experiment (see also Figure 5). The main experiment is then performed during the next $290\,\mathrm{ms}$ corresponding to the low vibration time window of the cryostat cycle. The rubidium atoms are cooled at site \textit{A} inside the MOT for $17\,\mathrm{ms}$ while the frequency conversion interface is in the LOCK mode. The interface is then switched to the QFC mode and $20\,\mathrm{ns}$ long write pulses are sent to the atomic memory. If a write photon is detected at detector D1, the atomic ensemble is read out after a DLCZ storage time $\tau_{A}$ by sending a $340\,\mathrm{ns}$ long read pulse. The emitted read photon is converted in the QFCDs and afterwards stored for $\tau_{B}$ in the AFC. During the storage, the $994\,\mathrm{nm}$ pump is gated off \cite{Maring2014} for $5\,\mu\mathrm{s}$ by the AOM behind the tapered amplifier in order to retrieve the read photon in a noiseless time window. The write/read process lasts for $1\,\mathrm{ms}$ until the Rb atoms are recaptured by a new MOT and the interface is switched to the lock mode for the next $17\,\mathrm{ms}$. After 15 MOT captures and the corresponding write/read trials, the sequence restarts at the next cryostat cycle, preparing a new AFC in the crystal.

\section{IV. Additional characterizations}
\label{sec:AddChar}

\subsection{Cross- and auto-correlation measurements}

To gain information about the non-classicality of the generated quantum states, we assess their normalized cross- and auto-correlation functions. In particular, we measure the normalized cross-correlation between the write and read photons 

\begin{equation}
g^{(2)}_{w,r}=\frac{p_{w,r}}{p_w \cdot p_r}
\end{equation}
where $p_{w,r}$ is the probability to detect a coincidence between the write and read photon, and $p_w$ ($p_r$) is the probability to detect a write (read) photon. Moreover, we can measure the heralded auto-correlation of a read photon 

\begin{equation}
\alpha = g^{(2)}_{r1,r2|w} = \frac{p_{r1,r2|w}}{p_{r1|w} \cdot p_{r2|w}}
\end{equation}
via a Hanbury Brown Twiss setup. Here, $p_{r1,r2|w}$ denotes the probability to measure a coincidence between both read photon detections conditioned on a write photon detection, and $p_{r1|w}$ ($p_{r2|w}$) is the probability to detect a read photon on the first (second) detector conditioned on a write photon detection.

The non-classicality of the correlations between write and converted read photons is assessed by the Cauchy-Schwarz inequality, which states that for classical fields the correlation function is bounded by

\begin{equation}
g^{(2)}_{w_A,r_B}\leq\sqrt{g^{(2)}_{w_A,w_A}g^{(2)}_{r_B,r_B}}
\end{equation}
where $g^{(2)}_{w_A,w_A}$ and $g^{(2)}_{r_B,r_B}$ are the unheralded auto-correlation functions of the write and read photons, respectively. The values that we measure to assess the inequality are shown in Table 2. The unheralded write photon autocorrelation shows a value of $g^{(2)}_{w_A,w_A}\approx2$ as expected for the ideal two-mode squeezed state generated by the atomic cloud. The unheralded autocorrelation function $g^{(2)}_{r_B,r_B}$ of the read photons at the output of the crystal was not taken because of unfeasible integration times. Instead, we measured its value at the output of the Rb cloud at site~\textit{A} and at the output of the second frequency conversion stage QFCD2 at site~\textit{B}. The values shown in Table 2 indicate that the read photons generated by the atomic cloud exhibit some bunching with  $1 \leq g^{(2)}_{r_A,r_A}\leq 2$. The values after the frequency conversion of $g^{(2)}_{r_\mathrm{QFCD2},r_\mathrm{QFCD2}}\approx 1$ indicate that the noise added by the QFCDs has Poissonian statistics. Hence, the read photon autocorrelation at the output of the crystal should not have a value higher than 2, indicating that $g^{(2)}_{w_A,r_B}>2$ is a strong sign of non-classicality.

\begin{table}[htb]
\caption{\textbf{Unheralded autocorrelation measurements.} Normalized autocorrelation values for the write and read photon fields. $g^{(2)}_{w_A,w_A}$ is measured after the write photons are filtered with the Fabry-Perot cavity. $g^{(2)}_{r_A,r_A}$ is measured at the output of the fiber that collects the read photons from the atomic cloud. $g^{(2)}_{r_\mathrm{QFCD2},r_\mathrm{QFCD2}}$ is measured at the output of QFCD2 at site~\textit{B}.}
\centering
\begin{tabular}{ K{1.2cm} | K{2cm} K{2cm}  K{2cm} }
	$p_\mathrm{e}\,[\%]$	& $g^{(2)}_{w_A,w_A}$ 	& $g^{(2)}_{r_A,r_A}$		& $g^{(2)}_{r_\mathrm{QFCD2},r_\mathrm{QFCD2}}$ \\ 	
	\hline	
	$35$					& $1.97(0.10)$	& $1.36(0.05)$		& $1.06(0.05)$ \\
	$10$					& $1.91(0.10)$	& $1.48(0.06)$		& $0.96(0.04)$ \\
	$5$						& $2.13(0.20)$	& $1.35(0.07)$		& $1.00(0.04)$ \\
\end{tabular}
\label{Table1}
\end{table}

\subsection{Weak coherent state conversion and storage}
\label{subsec:AddChar-WCS}
\begin{figure}
\includegraphics[width=1.0\columnwidth]{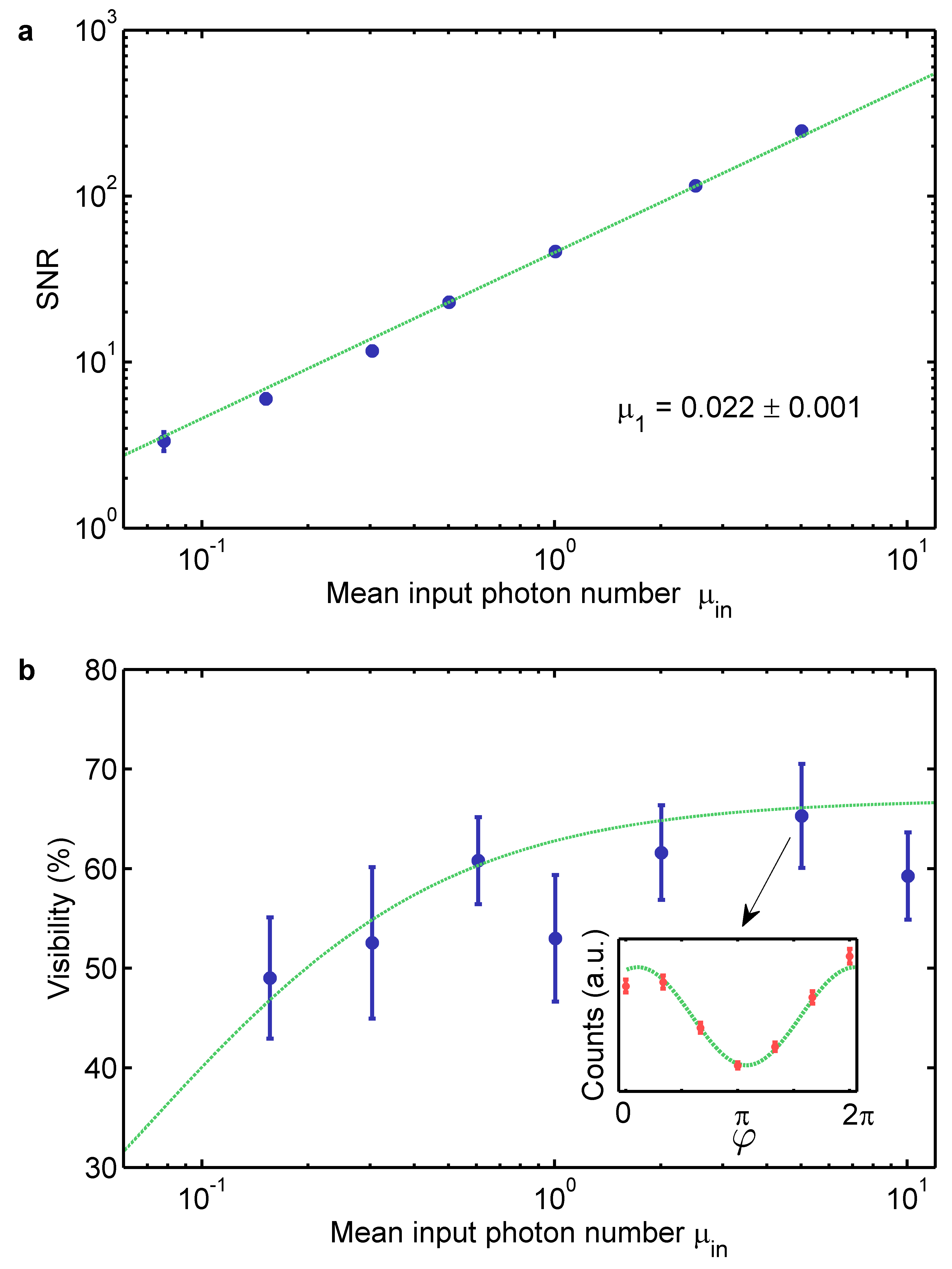}
\caption{\textbf{Weak coherent state measurements.} (a) SNR of the echo retrieved from the crystal, if a weak coherent state  is frequency converted in the QFCDs and stored in the memory depending on the mean input photon number per pulse $\mu_\mathrm{in}$ before the interface. The green line is a fit with the expected linear behaviour. (b) Visibility of interfering weak coherent time-bin pulses depending on their mean input photon number $\mu_\mathrm{in}$, after  the pulses were frequency converted and stored for $\tau_{B_1}=2\mu\mathrm{s}$ and $\tau_{B_2}=2.5\mu\mathrm{s}$ in the AFC memory. The green line is the predicted behavior of the visibility taking into account the measured SNR. The inset shows as an example the interference fringe taken at $\mu_\mathrm{in} \approx 5$.}
\label{WCS}
\end{figure}

The QFC interface, the storage in the crystal, and the locking system were first tested and characterized with weak coherent states of light mimicking the single read photons and time-bin qubits obtainable from the cold atomic QM. 

Using light from the $780\,\mathrm{nm}$ write laser which is sent through another AOM beam line and a set of neutral density filters (not shown in Figure 5), we generated attenuated laser pulses of Gaussian shape and $200\,\mathrm{ns}$ duration, at the same optical frequency of the read photons from the cold atomic QM. The weak laser pulses are converted through the QFCDs, then stored for $\tau_B=2.5\,\mu\mathrm{s}$ in the crystal and eventually retrieved and detected at D2. The obtained SNR of the retrieved echo is shown in Figure 10(a), as a function of the average input photon number per pulse. The linear fit highlights the performance in term of SNR of our system, showing $\mu_1 = 0.022 \pm 0.001$ {\textendash} with $\mu_1$ being the minimum number of photons per pulse at the input necessary to achieve a SNR of 1 for the detected echo. The echo then shows a SNR of $14$ for an average photon number of $0.3$ per pulse at $780\,\mathrm{nm}$ - corresponding to the expected number of read photons at the input of the first QFCD per heralded excitation in the cold atomic QM (fiber-coupled DLCZ retrieval efficiency $\eta_{A}^\mathrm{ret}\approx30\%$).

Next, weak coherent time-bin qubits -- attenuated doubly-peaked Gaussian pulses, separated by $500\,\mathrm{ns}$ and with tunable phase difference $\varphi$ between the early and late bins -- were sent through the QFCDs and the solid state storage device. The memory is prepared with two AFCs offering simultaneous storage for $\tau_{B_1}=2\mu\mathrm{s}$ and $\tau_{B_2}=2.5\mu\mathrm{s}$. The early and late bins are overlapped and the interference between the early and late pulses is measured as a function of the relative phase $\varphi$. The visibility of this interference is shown in Figure 10(b) as a function of the photon number per time-bin qubit $\mu_\mathrm{in}$. With strong coherent pulses the visibility of this interference is measured to be $V_0 = 67\%$. The decrease of visibility for lower input photon number $\mu_\mathrm{in}$ is due to a decrease in SNR. Taking this effect into account, the visibility becomes \cite{Gundogan2015}
\begin{equation}
V(\mu_\mathrm{in}) = V_0\frac{\mu_\mathrm{in}}{\mu_\mathrm{in}+2\beta\mu_{1}},
\end{equation}
where $V_0$ is the maximum visibility, and $\beta$ the correcting factor for the reduced efficiency of a double comb AFC compared to a single one. The simple model reproduces well our data, using the SNR measured in Figure 10(a) and $V_0$ measured with strong light pulses (see following section). The visibilities in the single photon regime, presented in the main text correspond here to a regime where $\mu_{in}\approx 0.3$ (mimicking the retrieval efficiency $\eta^B$ of the atomic memory).

\subsection{Fidelity limitation}
\label{subsec:AddChar-FidelityLimit}
In this section we discuss the limitations of the fidelity measured and presented in the main paper. 

The fidelity of the polar states is limited by the SNR of the detected photons: $\mathcal{F}_\mathrm{pol} \approx \frac{\mathrm{SNR}+1}{\mathrm{SNR}+2}$. The fidelity of the equatorial states is mainly limited by the visibility of the interference between early and late time-bins: $\mathcal{F}_\mathrm{eq} \approx (1+V)/2$. The visibility $\mathrm{V}$ depends on background noise (as shown in the previous section) and the overall frequency jitter of the lasers involved in the experiment.

Due to our relatively high SNR, the main limitation in our case is most likely given by laser jitter, which stochastically shifts the central frequency of the read photon. As seen in section~II a frequency shift $\delta$ induces a relative phase $2\pi\delta(\tau_1-\tau_2)$ between the two interferometer arms, thus reducing the measured visibility over several experimental trials.
All the lasers involved in the experiment contribute to this effect --- the $780\,\mathrm{nm}$ read laser generating the time-bin photon, the two pump lasers of the QFCDs converting the time-bin photon, the $780\,\mathrm{nm}$ write laser generating the lock light, and the $606\,\mathrm{nm}$ laser preparing the AFCs and acting as the reference for the beat-note lock.

Considering a Gaussian global laser linewidth of $\sigma$, the visibility $V_0$ of the interference between the two time-bins separated by $\Delta\tau$ can be expressed as \cite{Minar2008}
\begin{equation}\label{eqvisi}
V_0 = \mathrm{exp}\left(-\frac{\left(2\pi\sigma\Delta\tau\right)^2}{2}\right).
\end{equation}

\begin{figure}
\includegraphics[width=1.0\columnwidth]{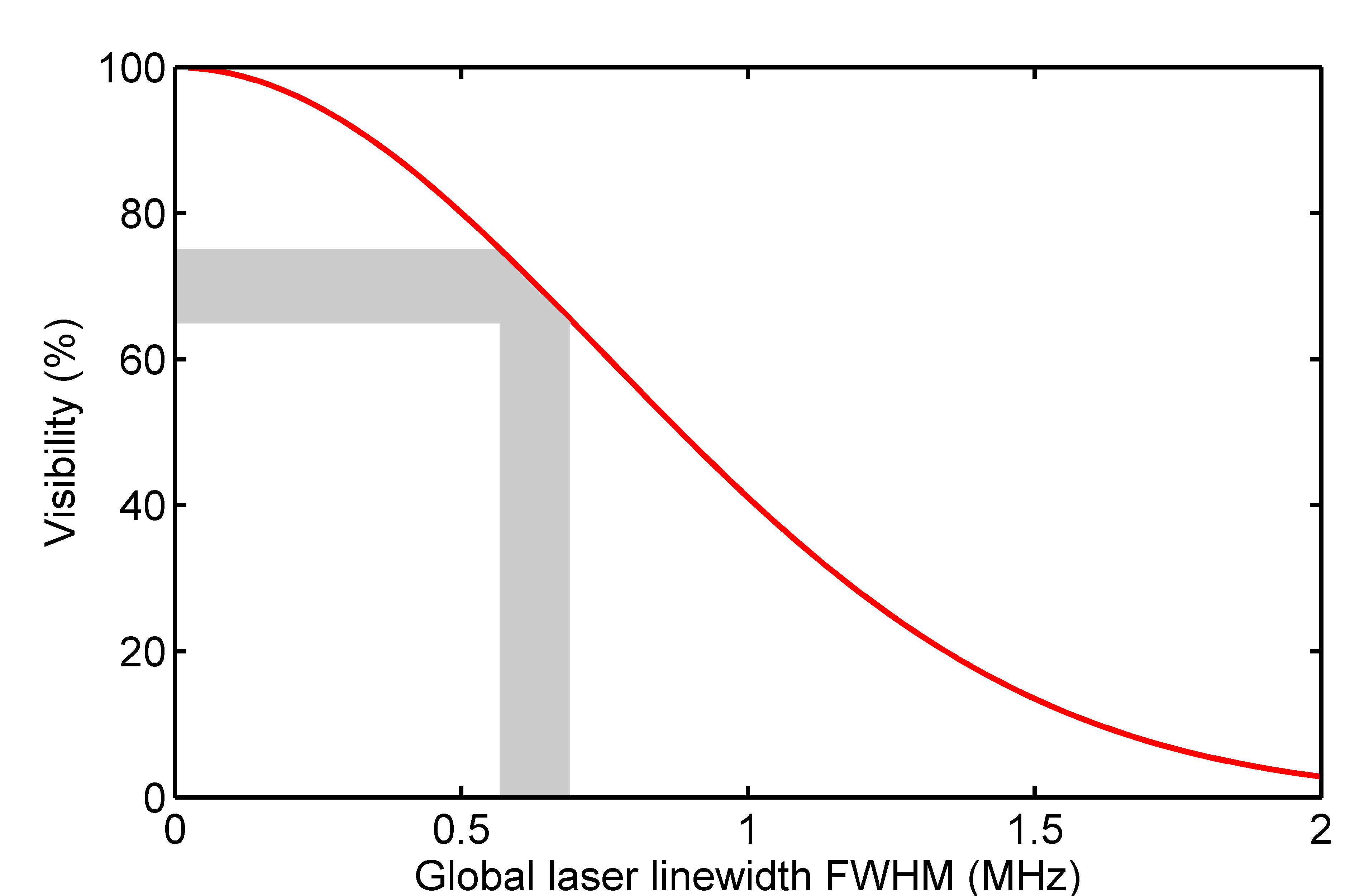}
\caption{\textbf{Interference visibility.} Visibility as a function of the laser linewidth FWHM ($2.35\,\sigma$) depicting Eq.~\ref{eqvisi}. The shaded area shows the operating range.}
\label{Visi}
\end{figure}

Depending on lasers stability, maximum visibilities $V_0$, measured with strong coherent pulses, between $65\%$ and $75\%$ have been observed. This corresponds to a global linewidth of the lasers between $570$ and $700\,\mathrm{kHz}$ FWHM ($\widehat{=}\,2.35\,\sigma$). Figure 11 shows the visibility as a function of the FWHM of the Gaussian linewidth, according to equation~(\ref{eqvisi}). When the pump laser at $994\,\mathrm{nm}$ was not stabilized on a Fabry-Perot cavity, we observed a significantly lower maximum visibility, around $60\%$ (corresponding to a visibility of around $50\,\%$ in the single photon regime).

\subsection{Storage time in the crystal}
\label{subsec:AddChar-StorageTime}

\begin{figure}
\includegraphics[width=1.0\columnwidth]{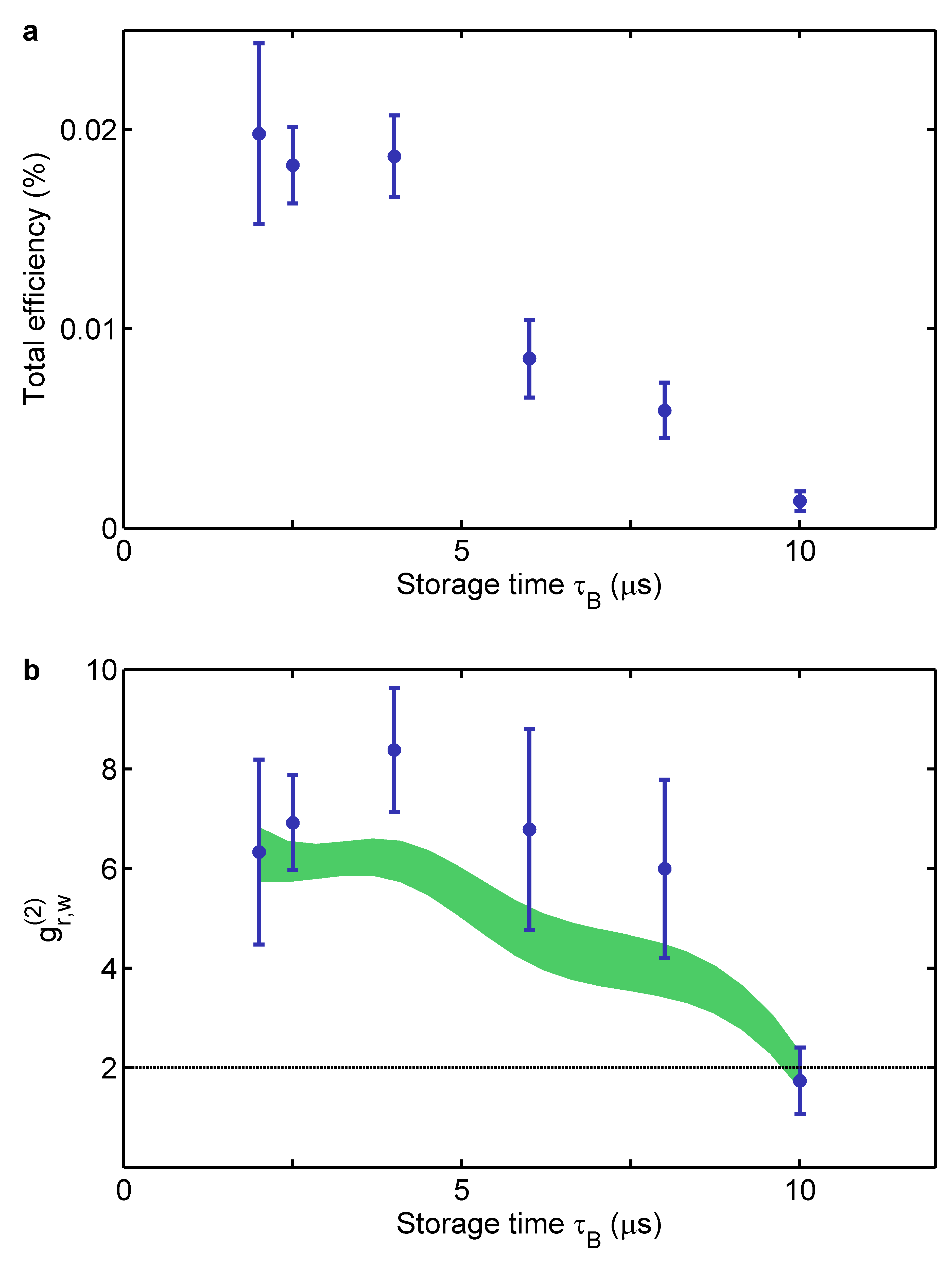}
\caption{\textbf{Storage efficiency and cross-correlation versus storage time.} Total storage efficiency and normalized cross-correlation function $g^{(2)}_{w,r_B}$ of the initial write photon at site \textit{A} and the converted, stored and retrieved read photon at site \textit{B} depending on the storage time $\tau_B$ in the crystal, taken at $p_\mathrm{e}\approx10\%$.}
\label{StorageTime}
\end{figure}
The preservation of non-classical correlations between the write photons -- detected at D1 at site \textit{A} -- and the converted, stored and retrieved read photons -- detected at D2 at site \textit{B} -- depending on the storage time $\tau_B$ in the crystal is investigated in Figure 12.

First, we show in Figure 12(a) the total detection probability at D2 of a converted-stored read photon when a heralding write photon is detected at D1. For small $\tau_B$ we obtain total efficiencies up to $0.02\%$ matching the expected range determined by individual optical losses. The decrease of the total efficiency over storage time follows the drop of the AFC memory efficiency due to the change in finesse and effective optical density of the prepared AFC at different $\tau_{B}$ caused by the finite laser linewidth \cite{Afzelius2009}. 

In Figure 12(b), the normalized cross-correlation $g^{(2)}_{w,r_B}$ between the write and read photons depending on $\tau_{B}$ is shown. We observe a relatively constant $g^{(2)}_{w,r_B}\approx 6$ up to a storage time of $\tau_B\approx 8\,\mu\mathrm{s}$ until it finally drops below the classical threshold of $g^{(2)}_{w,r}=2$ at $\tau_B\approx 10\,\mu\mathrm{s}$, where the AFC efficiency is low and its echo detection is limited by dark counts of the detector D2. The green area shows the expected correlations taking into account the SNR of the AFC echo, inferred from the AFC efficiencies of Figure 12(a), using the same model as \cite{Albrecht2014}.

\subsection{Data Availability}
The data that support the findings of this study are available from the corresponding author upon reasonable request.

} 

%


\bibliographystyle{osajnl}

%

%
%


\end{document}